\documentclass[conference]{IEEEtran}
\IEEEoverridecommandlockouts
\usepackage{cite}
\usepackage{my_sty}
\usepackage{amsfonts}
\usepackage{amsmath}
\usepackage{amstext}
\usepackage{amssymb}
\usepackage{algorithmic}
\usepackage{graphicx}
\usepackage{textcomp}
\usepackage{xcolor}
\usepackage{times}
\usepackage{epsfig}
\usepackage{booktabs}
\usepackage{multirow}
\usepackage{subfigure}
\usepackage{url}
\usepackage{subcaption}

\def\BibTeX{{\rm B\kern-.05em{\sc i\kern-.025em b}\kern-.08em
    T\kern-.1667em\lower.7ex\hbox{E}\kern-.125emX}}
    
\begin{document}

\title{CANF-VC++: Enhancing Conditional Augmented Normalizing Flows for Video Compression with Advanced Techniques}



\author{Peng-Yu Chen, Wen-Hsiao Peng \\
        \textit{National Yang Ming Chiao Tung University}\\
        wpeng@cs.nctu.edu.tw
}

\maketitle

\begin{abstract}
Video has become the predominant medium for information dissemination, driving the need for efficient video codecs. Recent advancements in learned video compression have shown promising results, surpassing traditional codecs in terms of coding efficiency. However, challenges remain in integrating fragmented techniques and incorporating new tools into existing codecs. In this paper, we comprehensively review the state-of-the-art CANF-VC codec and propose CANF-VC++, an enhanced version that addresses these challenges. We systematically explore architecture design, reference frame type, training procedure, and entropy coding efficiency, leading to substantial coding improvements. CANF-VC++ achieves significant Bjøntegaard-Delta rate savings on conventional datasets UVG, HEVC Class B and MCL-JCV, outperforming the baseline CANF-VC and even the H.266 reference software VTM. Our work demonstrates the potential of integrating advancements in video compression and serves as inspiration for future research in the field.
\end{abstract}

\section{Introduction}
\label{sec:intro}
Nowadays, video has emerged as the dominant form of conveying information due to its capacity to encompass a vast array of content, encompassing text, images, and sound. It is estimated that approximately 87\% of advertising services rely on video, while video data streams contribute to around 80\% of internet congestion~\cite{cisco}. Furthermore, there has been a surge in the resolution of video and screen size of multimedia devices. As a result, there is a demand for solutions that enhance or substitute traditional video codecs.        

Over the past few years, thanks to the development of artificial intelligence and high performance hardware devices, the field of end-to-end learned video compression has attracted significant research attention. The introduction of the first end-to-end learning-based video codec, namely deep video compression (DVC) model~\cite{dvclu}, in 2019 marked a major milestone. DVC is based on the architecture of a residual video coding framework. It utilizes neural network techniques for motion estimation, motion compensation, and coding of motion/residual signals within a hybrid video compression approach. Subsequent studies~\cite{ssf, mlvc, fvc, dcvc, canfvc, tcm, vct, acmmm22, mimt} have continued to push the boundaries, incorporating the latest advancements in deep learning to enhance the performance of different components in learned video compression.These studies have rapidly caught up with, and even surpassed, traditional video codecs like H.265/High Efficiency Video Coding (HEVC)~\cite{hevc} and H.266/Versatile Video Coding (VVC)~\cite{vvc} in terms of coding efficiency. However, amidst this rapid progress, a key challenge has emerged: effectively integrating techniques from different works in the field.

While individual studies have made significant progress in learned video compression, the fast-paced nature of development has resulted in fragmented techniques. Valuable insights and advancements from one study do not always transfer to others easily, hindering overall progress and collaboration within the community. This lack of integration poses a significant obstacle that needs to be overcome to fully leverage the collective potential of learned video compression.

One key challenge is the complexity of incorporating new tools into existing learned codecs. Training a learned video codec goes beyond simple end-to-end training; it often involves a process of progressively training each module~\cite{nvc, mlvc, canfvc}. This complexity makes it challenging and time-consuming to integrate new tools and techniques. The intricate nature of this process further impedes the widespread adoption and advancement of these algorithms.

To tackle these challenges, this work comprehensively reviews recent advancements in learned video compression. Our approach aims to enhance the performance of the state-of-the-art CANF-VC codec~\cite{canfvc} by carefully exploring various aspects. We systematically examine architecture design, reference frame formulation, training procedure, and entropy coding efficiency, conducting thorough tests to evaluate different variants and their impact on the baseline framework. Moreover, we propose an innovative approach to unleash the untapped potential of a specific training procedure, resulting in significant coding improvements beyond initial expectations.

The culmination of our efforts is the development of CANF-VC++, our final model. Through careful experimentation and the integration of various enhancements, CANF-VC++ demonstrates remarkable improvements, achieving substantial Bjøntegaard-Delta (BD) rate~\cite{bdrate} savings of 40.2\%, 38.1\%, and 35.5\% on widely used test benchmarks, including UVG~\cite{uvg}, HEVC Class B~\cite{hevcctc}, and MCL-JCV~\cite{mcl} datasets, respectively, when compared to the baseline framework CANF-VC~\cite{canfvc}. These results highlight the efficacy of our proposed approach and pave the way for further advancements in the field of learned video compression.
\section{Related Work}
\label{sec:related_work}

\subsection{Learned Video Compression}
\begin{figure*}
    \centering
    \subfigure[Traditional Codec]{
        \includegraphics[width=0.45\linewidth]{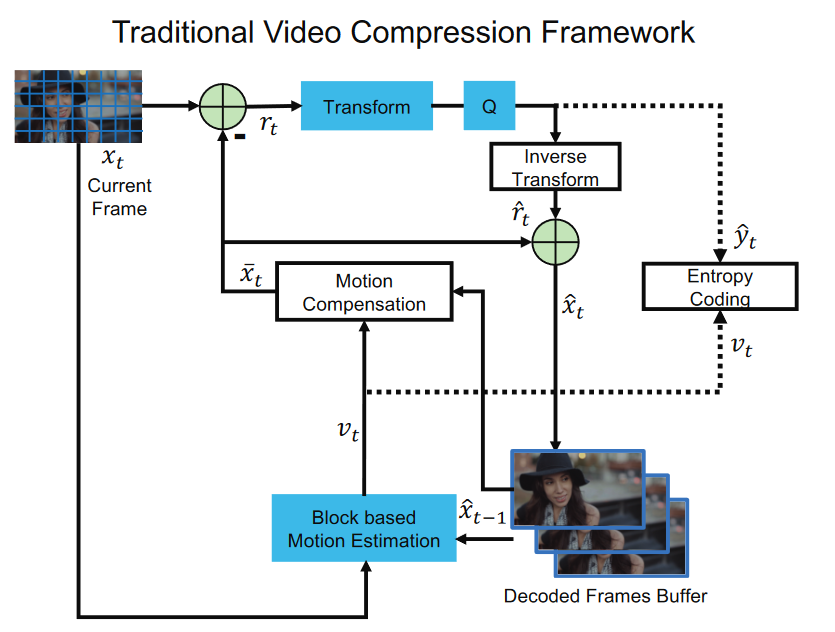}
        \label{fig:trad_codec}
    }
    \subfigure[DVC Codec]{
        \includegraphics[width=0.45\linewidth]{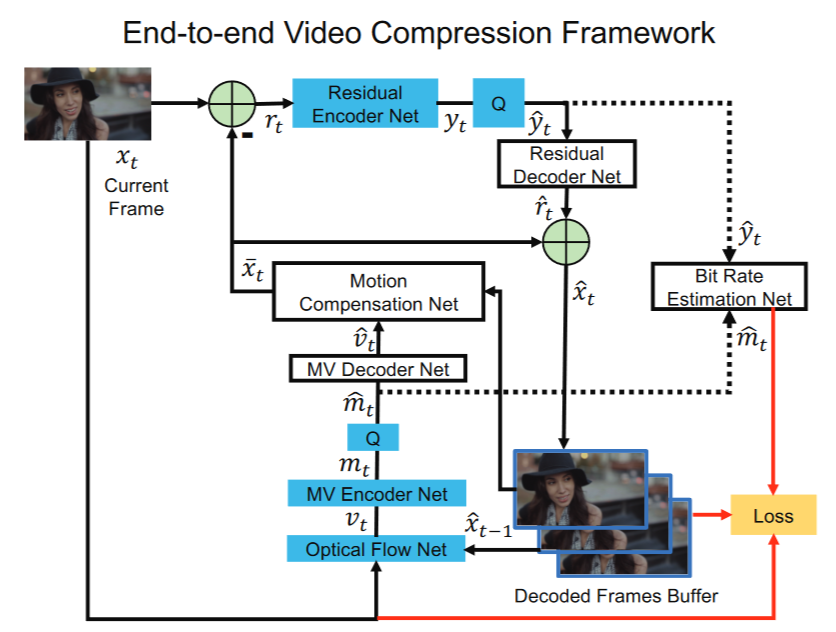}
        \label{fig:dvc_codec}
    }
    \caption{Comparison of (a) traditional video compression framework and (b) end-to-end hybrid-based coding framework~\cite{dvclu}. Source: Lu~\textit{et al.}~\cite{dvclu}.}
    \label{fig:dvc_compare}
\end{figure*}

Learned video compression has gained significant attention in the research community since the introduction of DVC~\cite{dvclu}. Overall, components of DVC model are one-to-one mapped from the architecture of residual coding and perform motion estimation, motion vector coding, motion compensation, residual coding by using neural networks within a hybrid-based coding framework. Figure~\ref{fig:dvc_compare} visually illustrates the profound impact of the DVC model's one-to-one mapping scheme, which directly adapts from the traditional hybrid-based coding framework. This mapping approach has become a foundational inspiration for subsequent works in learned video compression.

After that, many related studies were conducted and introduced various techniques to enhance different aspects of learned video compression. For example, Agustsson~\textit{et al.}~\cite{ssf} perform scale-space motion compensation to blur uncertain areas, such as fast motion and dis-occluded regions, in the frame predictor. Hu~\textit{et al.}~\cite{fvc} propose performing motion compensation and residual coding in the feature domain using deformable convolution~\cite{dai2017deformable}. Additionally, Lin~\textit{et al.}~\cite{mlvc} and Rippel~\textit{et al.}~\cite{elfvc} utilize a linear motion predictor to reduce motion coding overhead. These approaches aim to improve the temporal predictor and perform residual coding in either the pixel or feature domain.

\subsection{Conditional Coding for Learned Video Compression} 
Conditional coding has emerged as a fundamental paradigm in various state-of-the-art learned video compression systems~\cite{mmsp, dcvc, tcm, canfvc, acmmm22}. These systems consistently outperform conventional residual coding approaches~\cite{dvcpro, ssf, fvc, mlvc, elfvc} on average. The concept of conditional coding was introduced by Ladune~\textit{et al.}~\cite{mmsp} to capture the conditional entropy $H(x_t|x_c)$, where $x_t$ represents the video frame or feature to be encoded, and $x_c$ denotes the motion-compensated reference frame or feature. In contrast, residual coding aims to capture the residual entropy $H(r_t) = H(x_t - x_c)$. Theoretically, it has been demonstrated that the conditional entropy is inherently lower than or equal to the residual entropy, i.e., $H(x_t|x_c) \leq H(x_t - x_c)$, from an information-theoretic standpoint. This implies that learning the conditional entropy is more effective than learning the residual entropy.

Based on the theoretical analysis above, several studies have introduced video compression architecture using conditional coding.   
Li~\textit{et al.}~\cite{dcvc} expanded upon this concept by proposing a feature domain predictor for conditional inter-frame coding, exhibiting remarkable superiority over the pixel-domain residual coding framework~\cite{dvcpro}. Sheng~\textit{et al.}~\cite{tcm} further explored the utilization of multi-scale feature domain context for conditional inter-frame coding and introduced feature propagation (FP) as a long-term memory mechanism to augment coding efficiency. Their approach demonstrates coding efficiency on par with the H.266~\cite{vvc} reference model VTM~\cite{VVCSoftware_VTM} in Low-Delay-P configuration. Li~\textit{et al.}~\cite{acmmm22} enhanced the coding performance of \cite{tcm} by introducing a group-based context entropy model and a content-adaptive quantization mechanism. Ho~\textit{et al.}~\cite{canfvc} introduced a pure normalizing-flow-based conditional coding framework, extending the concept of conditional coding to optical flow map coding with a non-linear optical flow map predictor.

Mentzer~\textit{et al.}~\cite{vct} proposed an elegant conditional coding framework based on the Transformer architecture~\cite{vaswani2017attention}. Instead of encoding an optical flow map to construct a motion-compensated predictor, they individually compress frame latents and exploit the Transformer to model temporal dependencies by encoding previously decoded latents as conditioning signals. Xiang~\textit{et al.}~\cite{mimt} also introduced a Transformer model and employed the MaskGit~\cite{chang2022maskgit} technique for bi-directional autoregressive decoding. Their method offers a more flexible entropy coding scheme with superior coding performance. 

In conclusion, recent advancements in learned video compression consistently favor conditional coding over residual coding, highlighting its substantial potential when combined with advanced deep learning tools. Furthermore, conditional coding has been successfully applied in various learned video compression codecs, demonstrating high coding performance and paving the way for further improvements in the field.

\subsection{Conditional Augmented Normalizing Flows for Video Compression}

\begin{figure}
\centering
   \includegraphics[width=\linewidth]{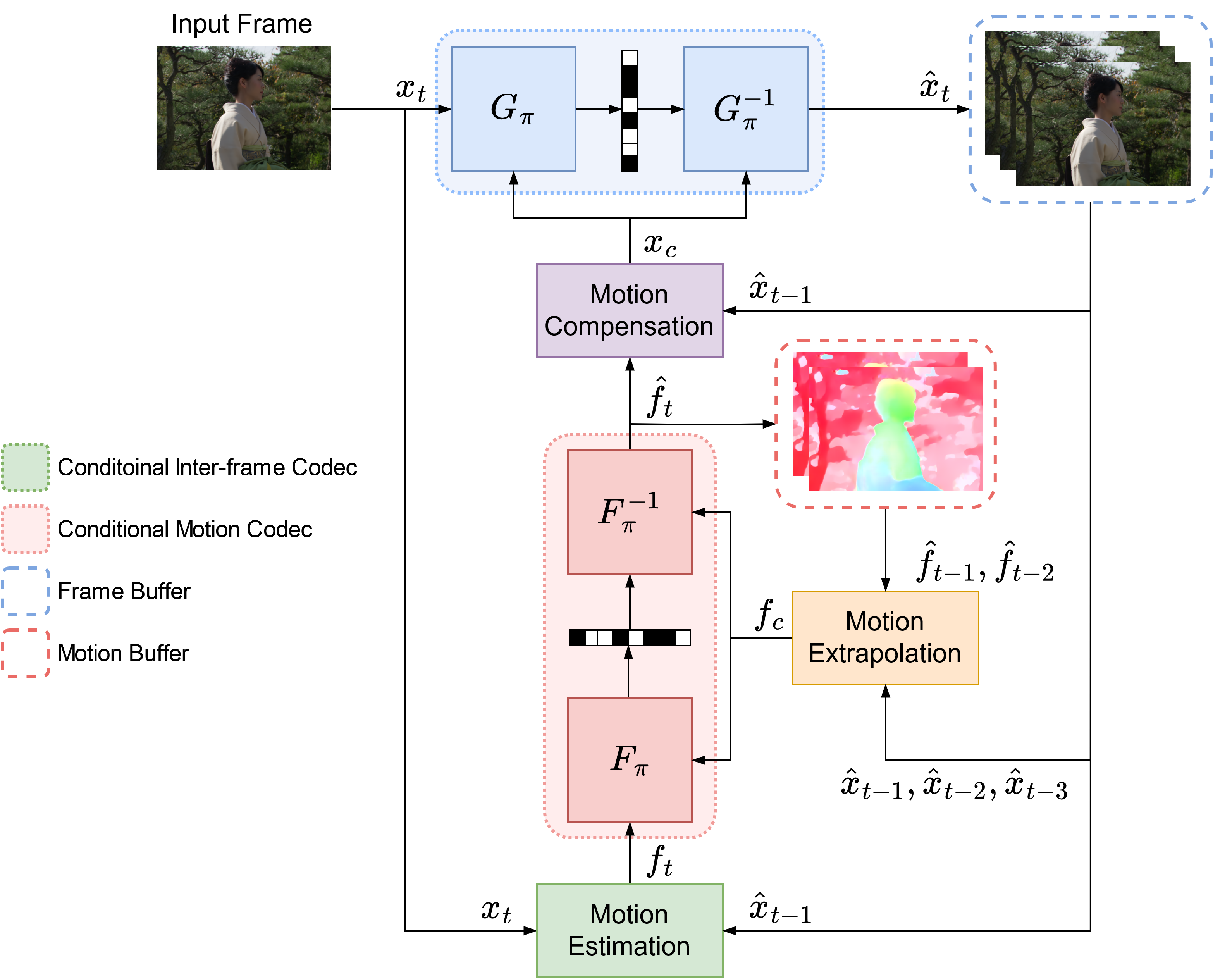}
   \caption{Architecture Overview of CANF-VC by Ho~\textit{et al.}~\cite{canfvc}. We adopt the architecture as our baseline framework.}
   \label{fig:system_overview_canfvc}
\end{figure}

In the field of conditional coding for learned video compression, Ho~\textit{et al.}\cite{canfvc} introduced a novel framework called Conditional Augmented Normalizing Flow (CANF) for pure conditional coding. Figure~\ref{fig:system_overview_canfvc} presents an overview of the learned P-frame framework proposed by Ho~\textit{et al.}\cite{canfvc}. The framework involves encoding a video sequence $\{x_i | i \in [1, T]\}$ consisting of $T$ frames. The first frame, $x_1$, is encoded as an I-frame using a method introduced by Ho~\textit{et al.}\cite{anfic}. The remaining frames, $\{x_i | i \in [2, T]\}$, are encoded using a P-frame coding scheme.

To encode frame $x_t$ at time stamp $t$, the process entails two conditional coding steps: conditional motion coding and conditional inter-frame coding.

For conditional motion coding, the optical flow $f_t$ is initially estimated using PWC-Net~\cite{pwc} and compressed using a CANF-based codec. To effectively encode the optical flow $f_t$, a conditional motion coding scheme is employed. As part of this scheme, an extrapolated optical flow map $f_c$ is generated by utilizing three decoded frames ($\hat{x}_{t-1}$, $\hat{x}_{t-2}$, $\hat{x}_{t-3}$) and two decoded optical flow maps ($\hat{f}_{t-1}$, $\hat{f}_{t-2}$). The extrapolated optical flow map $f_c$ is then employed as a conditioning signal, thereby reducing the motion coding overhead. In addition, intra motion coding is applied to the first P-frame using a variational autoencoder (VAE) with mean-scale hyperprior~\cite{googleiclr18} to compress $\hat{f}_2$.

For conditional inter-frame coding, the inter-frame prediction $x_c$ is generated by aligning the reference frame $\hat{x}_{t-1}$ with the decoded optical flow map $\hat{f}_t$. In the approach proposed by Ho~\textit{et al.}~\cite{canfvc}, a bilinear warping operation is employed on $\hat{x}_{t-1}$ based on the decoded optical flow map $\hat{f}_t$. Following this, the resulting warped frame is refined using a motion compensation network to produce the inter-frame prediction $x_c$. Subsequently, the inter-frame coding can be performed to encode $x_t$, utilizing $x_c$ as the conditioning signal. The final reconstructed frame $\hat{x}_t$ is obtained and buffered for encoding subsequent frames.

Overall, the Conditional Augmented Normalizing Flows for Video Compression (CANF-VC) framework presented by Ho~\textit{et al.}~\cite{canfvc} offers a unique approach for conditional coding in learned video compression. By incorporating both conditional motion coding and conditional inter-frame coding, their framework demonstrates the potential for improved compression performance in video sequences.

\subsection{Training Protocol of Learned Video Compression} 
In the context of learned video compression, the training procedure plays a crucial role in determining the overall rate-distortion performance. However, as highlighted by Liu~\textit{et al.}~\cite{nvc}, the strong dependency between inter-frame prediction coding and residue information coding within the hybrid-based coding framework commonly adopted by learned video codecs poses a challenge for achieving end-to-end optimization. This interdependency adds complexity to the optimization process and hinders the effective optimization of the entire codec. To address this, Liu~\textit{et al.}~\cite{nvc} propose a progressive training approach, which involves training the submodules in the learned codec progressively. This sequential training process includes motion estimation network training, motion codec training, motion compensation network training, and inter-coding model training before training the entire learned codec end-to-end. Additionally, the number of frames in the training sequences is progressively increased during the training process, ensuring stability and robustness.

Furthermore, various works have aimed to enhance coding efficiency by aligning the training procedure with the inference stage. Lu~\textit{et al.}\cite{Lu2020ContentAA} propose error propagation aware (EPA) training to mitigate error accumulation in video coding. They achieve this by jointly optimizing the entire training sequence, allowing the gradient to backpropagate through frames of different time stamps. Other approaches, such as those implemented in~\cite{ssf, tcm, acmmm22, li2023neural}, utilize a round-based training method inspired by Minnen and Singh's work on image compression~\cite{minnen2020channelwise}. In round-based training, actual rounding operations are applied to the latent variables whenever the latent variables are transformed through subsequent network, with a straight-through estimator used to handle the gradient flow during backpropagation.

Additionally, several methods~\cite{elfvc, mentzer2022neural, guo2023learning} propose applying different weighting on the distortion loss term for different P-frames during training. By using monotonically increasing weights to modulate the distortion loss, these approaches aim to mitigate drift errors or error accumulation in learned P-frame codecs.

In conclusion, the training protocol in learned video compression plays a vital role in achieving optimal rate-distortion performance. Progressive training approaches, such as the one proposed by Liu~\textit{et al.}~\cite{nvc}, and techniques that align the training process with the inference stage, like EPA training and round-based training, have shown promise in improving the effectiveness and efficiency of learned video codecs. Additionally, methods that employ weighted distortion loss for different P-frames during training help address issues related to drift errors and error accumulation. These advancements in training protocols contribute to the continued progress and development of learned video compression techniques.

\section{Proposed Method}
\label{sec:proposed_method}

\begin{figure}
   \includegraphics[width=\linewidth]{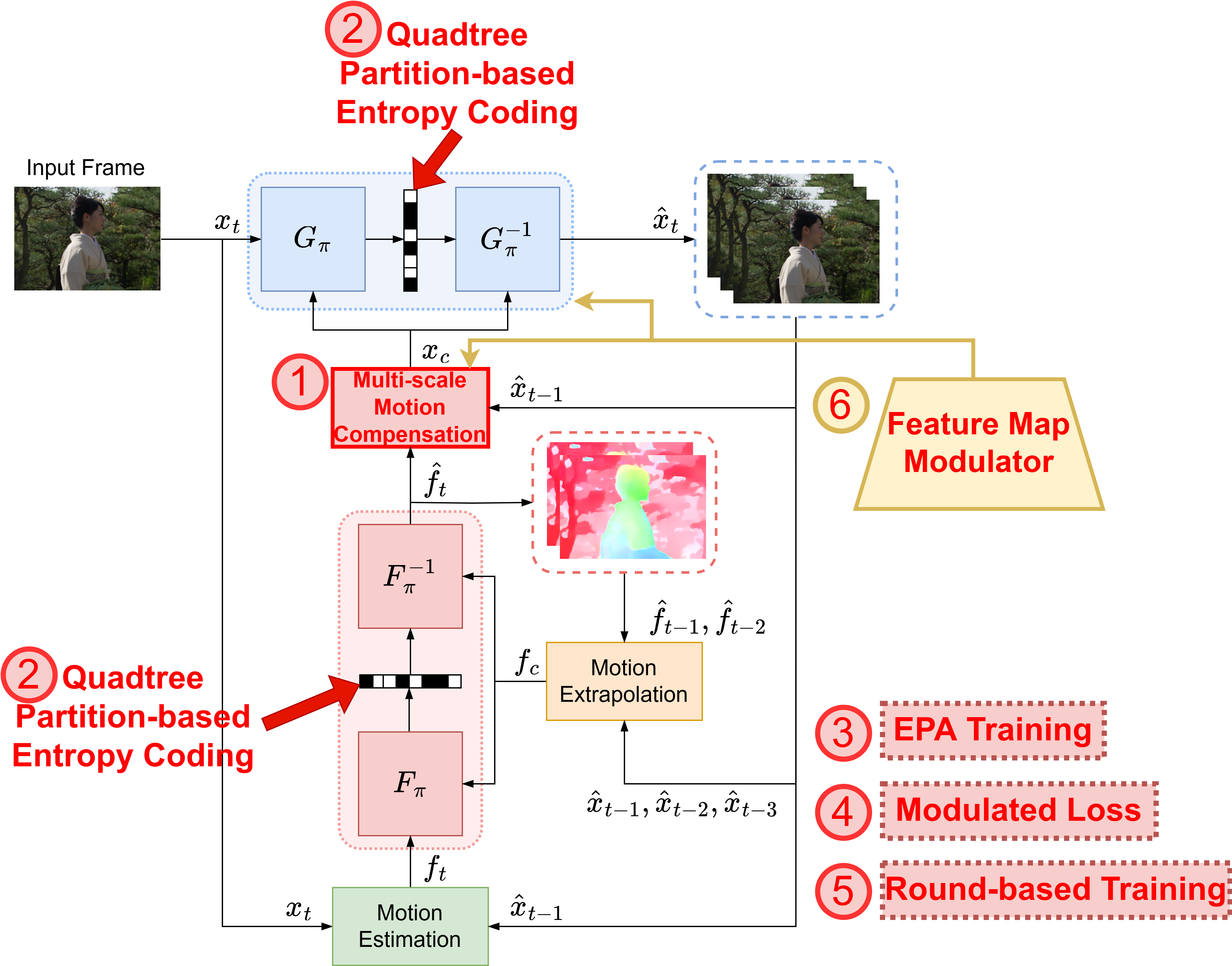}
   \caption{Architecture Overview of the proposed CANF-VC++ with improvements from baseline framework.We integrate several tools from other works in learned image/video compression, marked from `1` through `5`. We also propose a novel feature map modulator, marked as `6`, to incorporate with modulated loss (marked as `5`). The feature map modulator modulates intermediate feature maps of both multi-scale motion compensation network and inter codec $G_\pi, G_\pi^{-1}$. Please refer to Section~\ref{ssec:drift_error} for more details.}
   \label{fig:system_overview_canfvcpp}
\end{figure}

In our research, our main goal is to enhance the performance of the baseline framework proposed by Ho et al.~\cite{canfvc}. To achieve this objective, we are introducing a set of additional tools in this chapter, which will be integrated with our proposed enhancements. Then, we provide implementation details for effectively including these improvements.

\begin{figure*}
    \centering
    \subfigure[Motion compensation network utilized by Ho~\textit{et al.}~\cite{canfvc}.]{
      \includegraphics[width=0.48\linewidth]{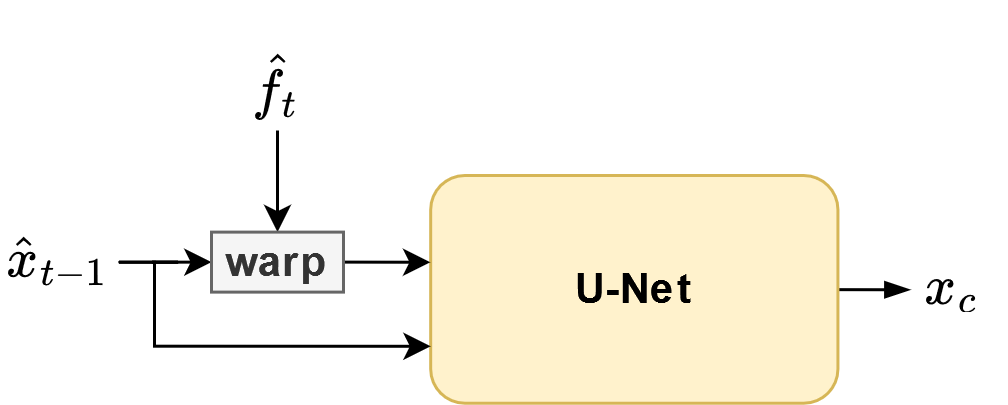}
      \label{fig:mcnet}
    }
    \subfigure[Multi-scale motion compensation network inspired by Chen~\textit{et al.}~\cite{chen2022bcanf}. The "warp \& concat" operation stands for performing bilinear warping using $\hat f_t$ and concatenate with the original, zero-motion input.]{
      \includegraphics[width=0.48\linewidth]{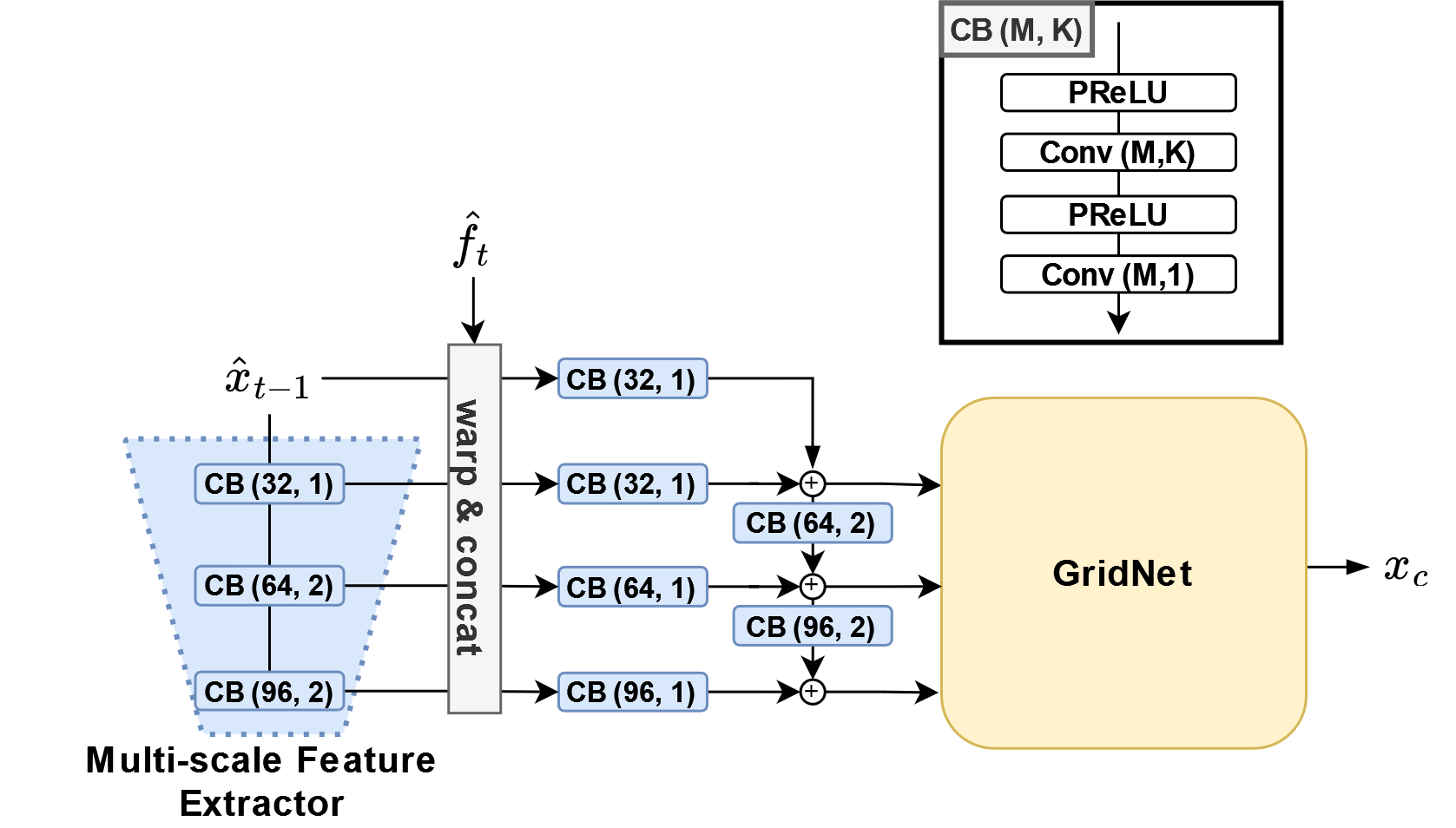}
      \label{fig:gridnet}
    }

    \subfigure[Our implementation of GridNet~\cite{fourure2017residual} in (b). We follow the design of GridNet~\cite{fourure2017residual} to build the network with residual block (green units), convolutional layers with strides (red units), and upsampling layers with convolutional layers (yellow units). A zoom on the dashed red square part with a detailed composition of each block is shown in (d). Source: Fourure~\textit{et al.}~\cite{fourure2017residual}.]{
      \includegraphics[width=0.48\linewidth]{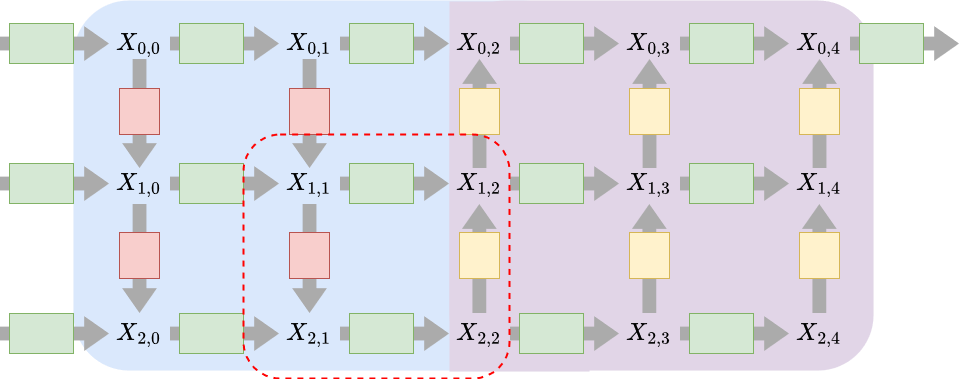}
      \label{fig:gridnet_detail}
    }
      
    \subfigure[Detailed schema of a GridBlock. Note that we remove the batch normalization layers and replace activation functions that used in GridNet~\cite{fourure2017residual}. Source: Fourure~\textit{et al.}~\cite{fourure2017residual}.]{
      \includegraphics[width=0.48\linewidth]{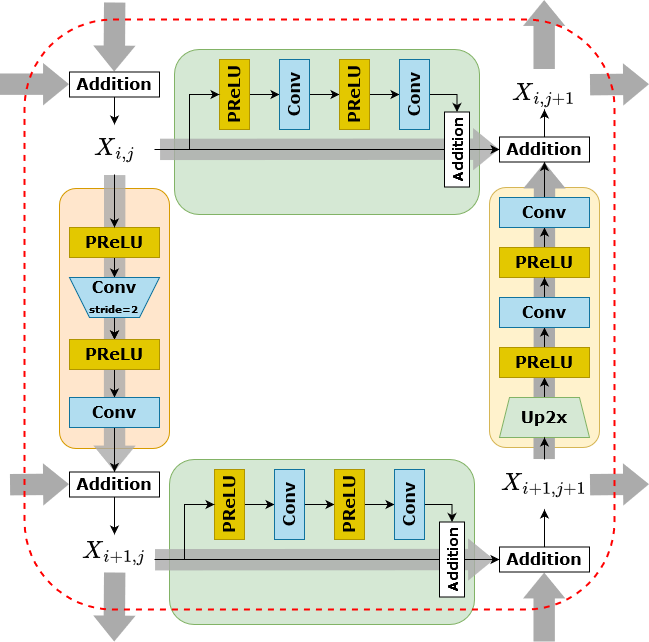}
    \label{fig:gridblock}
    }

    \caption{Illustration of (a) U-Net~\cite{ronneberger2015unet} based motion compensation network in Ho~\textit{et al.}~\cite{canfvc} and (b) our improved multi-scale motion compensation network based on GridNet~\cite{fourure2017residual} structure. (c)(d) The detailed architecture of our GridNet implementation in (b).}
    \label{fig:comparison_mcnet}
  
\end{figure*}

\subsection{Coding Tools Integration}
In order to further enhance the performance of the baseline framework proposed by Ho~\textit{et al.}~\cite{canfvc}, we have identified potential areas for improvement. We improve the baseline framework in several aspects, including conditioning signal generation, entropy model improvement, training tricks integration, and drift error mitigation. By introducing new techniques, we aim to overcome limitations and achieve significant advancements in video compression.

\subsection{Conditioning Signal Generation}
We have identified that the formulation of the conditioning signal in the baseline framework is overly simplistic. In the baseline framework, the conditioning signal $x_c$ for conditional inter-frame coding is generated by enhancing the motion-compensated reference frame using a simple U-Net~\cite{ronneberger2015unet}, as depicted in Figure~\ref{fig:mcnet}. Recent works have explored various aspects to improve inter-frame prediction efficiency, including multi-hypothesis prediction~\cite{mlvc, feng2021versatile, fvc, guo2023learning, li2023neural}, feature domain inter-frame prediction~\cite{fvc, dcvc, tcm}, and multi-scale context mining~\cite{nvc, guo2023learning, tcm, acmmm22, li2023neural}. These variants offer opportunities for improvement.

\textbf{Improvement: Multi-scale Motion Compensation}

To leverage the benefits of performing inter-frame prediction in the feature domain~\cite{fvc, dcvc} and multi-scale approaches~\cite{tcm, acmmm22, li2023neural}, we adopt a multi-scale motion compensation scheme inspired by the B-frame coding method proposed by Chen \textit{et al.}~\cite{chen2022bcanf}. We extract multi-scale features from the reference frame $\hat{x}_{t-1}$ and employ GridNet~\cite{fourure2017residual} as our multi-scale MCNet. This approach, illustrated in Figure~\ref{fig:gridnet}, enables us to generate a 3-channel motion-compensated signal $x_c$ for conditional inter-frame coding. We utilize the decoded optical flow map $\hat{f}_t$ to perform motion compensation on the multi-scale features, along with $\hat{x}_{t-1}$, allowing for down-sampling and rescaling of $\hat{f}_t$ if required. The motion-compensated features/frames are then concatenated with their zero-motion counterparts for GridNet forwarding, resulting in the conditioned signal $x_c$. This improvement in conditioning signal generation provides a more effective representation for conditional inter-frame coding, capitalizing on the advantages of multi-scale motion compensation and feature exploitation.


\subsection{Drift Error Mitigation}
\label{ssec:drift_error}
One of the improvements to be made in the baseline framework is techniques to handle drift error, which is commonly observed in P-frame coding schemes. Drift error refers to the accumulation of frame reconstruction errors over time, leading to a degradation in coding efficiency. Recent studies~\cite{elfvc, mentzer2022neural, guo2023learning} have introduced the concept of \textit{modulated loss}, which applies a monotonically increasing weighting on the distortion loss for encoding different P-frames during training. This approach guides the codec to reconstruct frames with higher quality to counteract the effect of drift error to some extent.

\textbf{Improvement: Modulated Loss with Feature Map Modulation}
To address the issue of drift error observed in P-frame coding schemes, we propose an improvement in the baseline framework by incorporating the concept of modulated loss. Modulated loss, introduced in recent studies~\cite{elfvc, mentzer2022neural, guo2023learning}, applies a monotonically increasing weighting on the distortion loss for encoding later P-frames during training to counteract the accumulation of reconstruction errors over time.

In our proposed method, we integrate the strictly increasing modulated loss, as proposed by Guo~\textit{et al.}~\cite{guo2023learning}, into the original training objective of the baseline framework proposed by Ho~\textit{et al.}~\cite{canfvc}. The modified training objective includes additional terms that facilitate the application of modulated loss, as shown below:
\begin{equation}
\begin{split}
\mathcal{L}_t^{mod} = \lambda \cdot \mu_t \cdot D_t + R_t + L_t^{reg}
\label{eq:oveall_training_objective_mod_loss}
\end{split}
\end{equation}

These terms represent the bit-rate estimation $R_t$ for encoding frame $t$, the distortion measurement ($D_t$), the modulation parameter $\mu_t$, and and $L_{reg}$ denotes regularization terms proposed in Ho~\textit{et al.}~\cite{canfvc}. The modulation parameter $\mu_t$ is introduced to control the weighting of the distortion term, with a monotonically increasing value for each P-frame. We adopt the modulated loss in Guo~\textit{et al.}~\cite{guo2023learning} and set $\mu_t$ as $\mu_2 = 1$ and $\mu_{t+1} = \mu_t + 0.2$.
 
\begin{figure}[t]
    \begin{center}
    \includegraphics[width=0.5\linewidth]{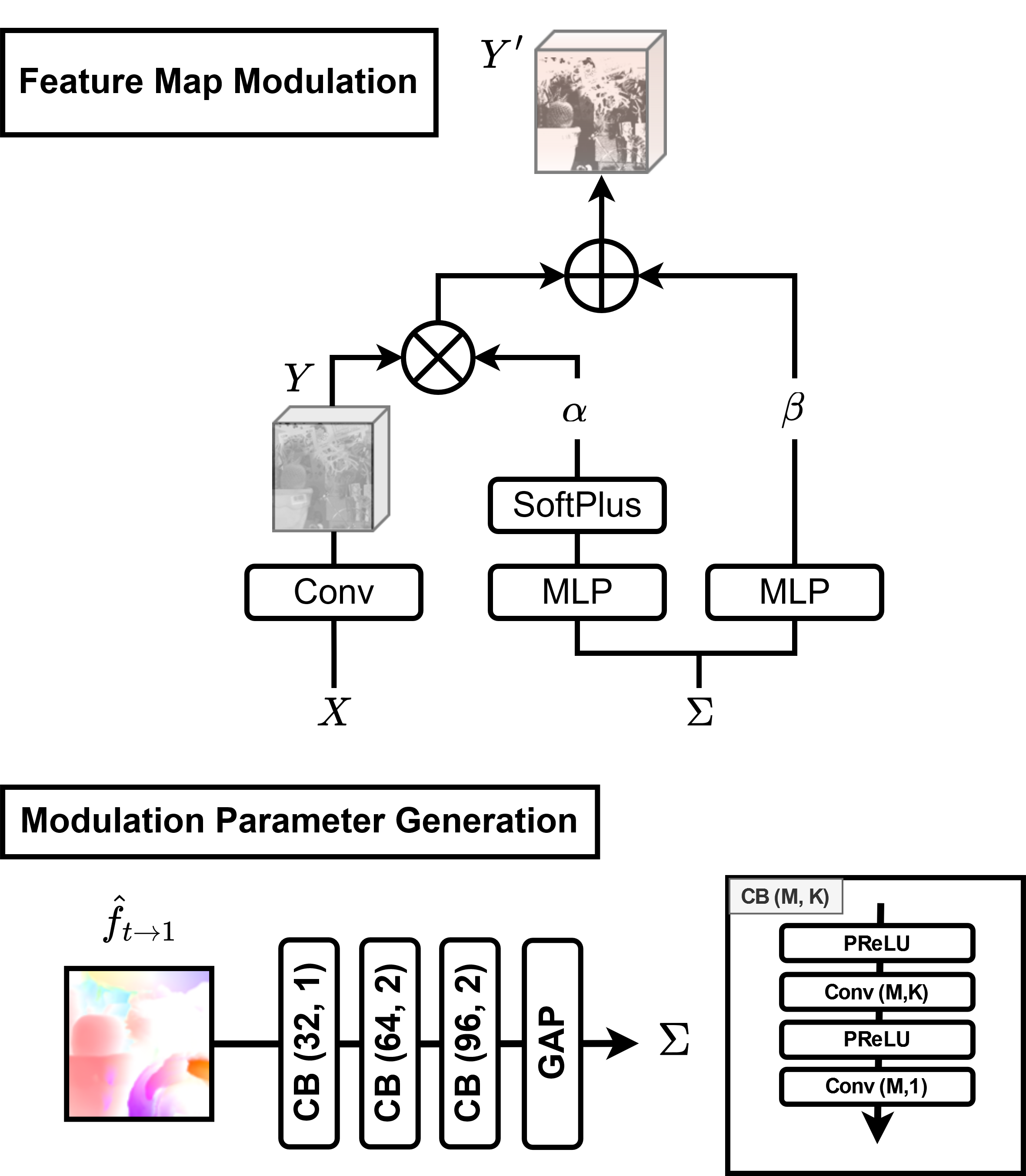}
    \end{center}
    \caption{Illustration of the feature map modulation to incorporate with modulated loss.}
    \label{fig:MCFT}
\end{figure}

Notably, we introduce a mechanism to enhance the baseline framework and effectively incorporate the modulated loss. To achieve this, we introduce a set of additional parameters that are integrated into the process of multi-scale motion compensation and conditional inter-frame coding, as depicted in Figure~\ref{fig:system_overview_canfvcpp}. The objective of these modulation parameters is to selectively enhance or suppress feature maps in order to improve the coding performance.

To begin with, we extract a globally accessible 1-D feature vector, denoted as $\Sigma$, from the temporal-propagated optical flow map $\hat f_{t \to 1}$. This optical flow map $\hat f_{t \to 1}$ is updated following the decoding of the optical flow map at each time index $t$ using the following rule:
\begin{equation}
    \hat f_{t \to 1} = 
    \begin{cases}
        \hat f_t, & \text{if } t=2, \\
        Warp(\hat f_{t-1 \to 1}, \hat f_t) + \hat f_t, & \text{otherwise}.
    \end{cases}
\end{equation}

The rationale behind using the temporal-propagated optical flow map $\hat f_{t \to 1}$ lies in its ability to estimate the optical flow between the current frame $x_t$ and the first frame $\hat x_1$ in a Group-of-Pictures (GOP). As $t$ increases, the lengths of optical flows in $\hat f_{t \to 1}$ also increase. This characteristic aligns well with the modulated loss, which monotonically increases with $t$. Despite $\hat f_{t \to 1}$ being a rough estimation due to the accumulation of compensation errors over time, we utilize it as a special state signal to extract the 1-D feature vector $\Sigma$. The spatial values of this vector are averaged to minimize any adverse impact on the optical flow quality.

Subsequently, we generate a pair of 1D vectors, $\alpha$ and $\beta$, from $\Sigma$, which serve as channel-wise descriptors. These descriptors are used to adapt the feature maps through the following transformation:

\begin{equation}
Y' = \alpha \otimes \text{Conv}(X) \oplus \beta,
\end{equation}

where $X$ represents the input, and $\otimes$ and $\oplus$ denote channel-wise multiplication and addition, respectively. The resulting feature map is denoted as $Y'$. {We apply feature map modulation on both multi-scale motion compensation network and inter codec $G_\pi, G_\pi^{-1}$, as demonstrated in Figure~\ref{fig:system_overview_canfvcpp}. By applying this modulation, our codec learns to enhance or suppress specific features based on the channel-wise descriptors $\alpha$ and $\beta$, thereby improving the overall rate-distortion performance.

It is important to note that while our original intention was to design an adaptive feature modulation technique that responds to different video contents, extensive experiments~\ref{exp:ablation_state_signal} have revealed that this mechanism does not exhibit significant sensitivity to different observing signals during testing. In other words, the performance of the codec is not significantly affected regardless of how the temporal-propagated optical flow map $\hat f_{t \to 1}$ changes. Despite this deviation from our initial design motivation, the incorporation of this mechanism, categorized as a tool to work in conjunction with the modulated loss, significantly enhances the coding performance and improves the overall rate-distortion performance of the codec.

\subsection{Entropy Coding Efficiency}
An area that requires improvement in the baseline framework is the selection of an entropy model. While Ho~\textit{et al.}~\cite{canfvc} rely solely on entropy models that utilize temporal information and side information for entropy estimation, it is crucial to acknowledge the existence of various approaches that go beyond a hyperprior model. These works have demonstrated improvements in compression performance by explicitly capturing either spatial dependencies~\cite{minnen2018, cheng2020, he2021checkerboard}, channel dependencies~\cite{minnen2020channelwise}, or even both~\cite{he2022elic, acmmm22, li2023neural}, leading to demonstrated improvements in compression performance.

\textbf{Improvement: Quadtree Partition-Based Entropy Model}
 
\begin{figure}[t]
    \begin{center}
    \includegraphics[width=\linewidth]{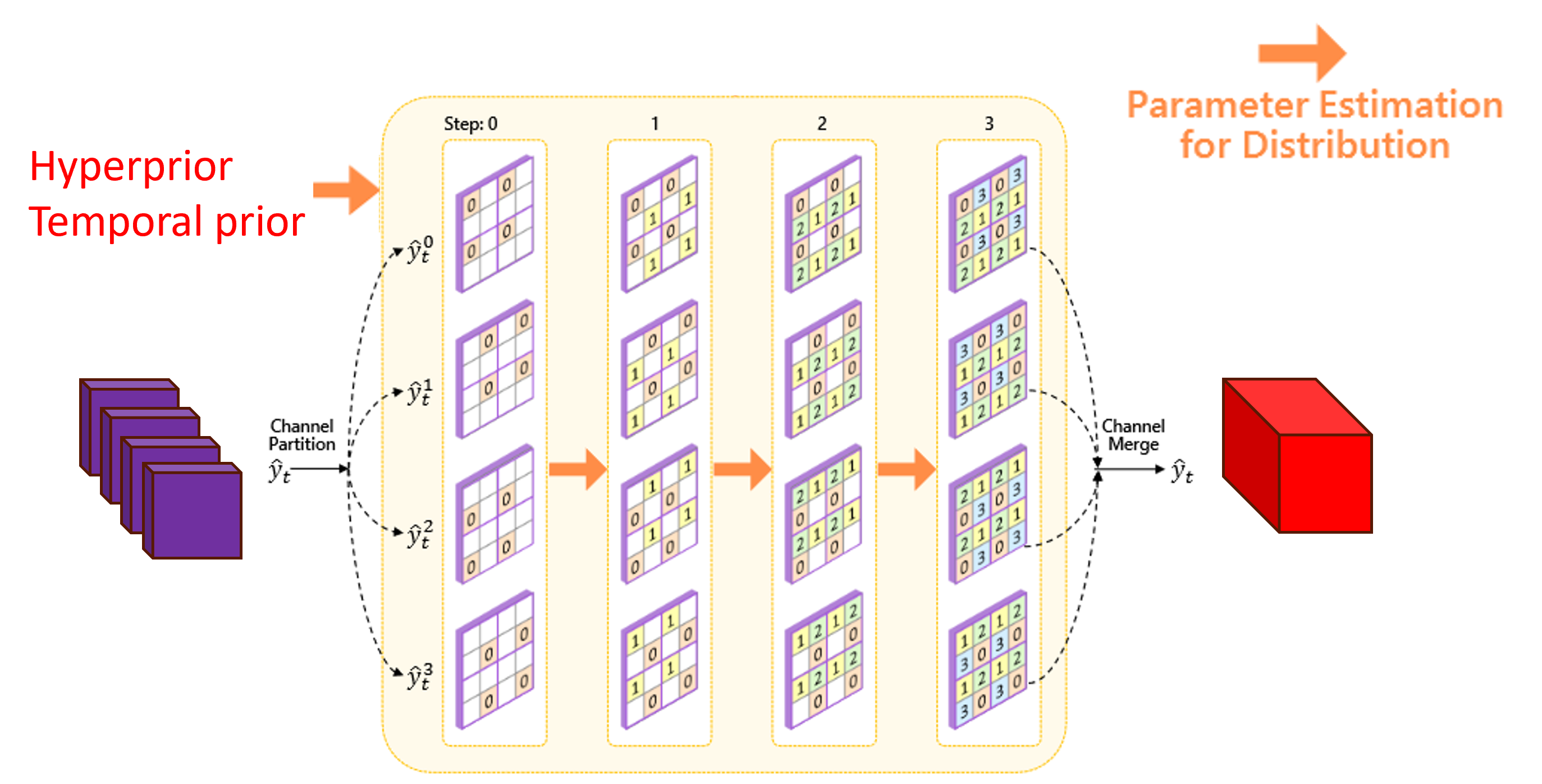}
    \end{center}
    \caption{Illustration of the quadtree partition-based entropy model proposed by Li~\textit{et al.}~\cite{li2023neural}. Source: Li~\textit{et al.}~\cite{li2023neural}.}
    \label{fig:context}
\end{figure}

To overcome the limitations of the baseline video compression framework and further enhance entropy coding efficiency, we introduce the quadtree partition-based entropy model proposed by Li~\textit{et al.}~\cite{li2023neural}. This model offers an effective approach for video compression by capturing both spatial and channel dependencies in the latent video data. In Li~\textit{et al.}~\cite{li2023neural}, they apply this quadtree partition-based entropy model for intra-frame coding, motion coding and conditional inter-frame coding to improve coding efficiency. The architecture of quadtree partition-based entropy model is shown in Figure~\ref{fig:context}.

As illustrated in Figure~\ref{fig:context}, the quadtree partition-based entropy model comprises two essential components. First, the latent $\hat y_t$ is divided into four groups along the channel dimension, ensuring an equitable distribution of channels across the groups. This division facilitates the modeling of channel dependencies.

Second, within each group, the video frames are divided into non-overlapping 2$\times$2 patches, resulting in four spatial parts for each group, marked as $0,1,2,3$ in Figure~\ref{fig:context}. By interleaving these spatial parts, the model effectively creates four groups along the spatial-channel dimension. This approach captures spatial dependencies within each group, thereby enhancing compression performance.

By combining the division along the channel dimension with the spatial partitioning, the quadtree partition-based entropy model maximizes the utilization of both spatial and channel contexts. This approach enables the model to accurately represent the latent video data by capturing dependencies at different scales and across multiple dimensions.

Compared to the checkboard-shaped partition model proposed by He~\textit{et al.}~\cite{he2021checkerboard}, which focuses on a specific spatial context, and other methods that solely address either spatial or channel dependencies, the quadtree partition-based entropy model offers a more comprehensive and flexible solution. It simultaneously addresses both spatial and channel dependencies, making it a versatile choice for video compression tasks. The model harnesses the benefits of its general context modeling capabilities, its ability to capture spatial and channel dependencies concurrently, and its suitability for real-time processing. These qualities validate our choice to introduce this model as an improvement over the baseline framework, and we anticipate significant enhancements in compression performance for practical video compression scenarios.

\subsection{Training/Testing Misalignment}
The baseline model suffers from a deficiency in aligning the codec behavior between the training and testing stages. To address this, we aim to review and introduce techniques that improve the coding performance by achieving better alignment.

\textbf{Improvement: Error Propagation Aware (EPA) Training}

\begin{figure*}
    \centering
    \subfigure[Training of DVC~\cite{dvclu} or Ho~\textit{et al.}~\cite{canfvc}.]{
        \includegraphics[width=0.45\linewidth]{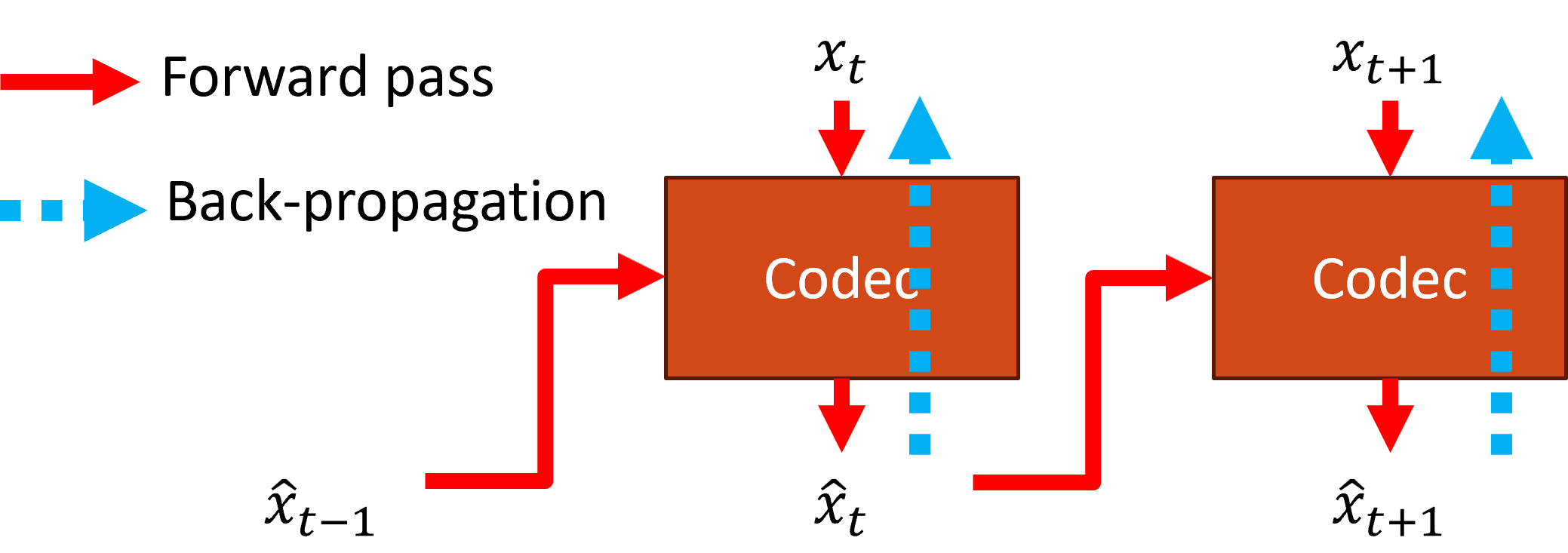}
        \label{fig:wo_EPA}
    }
    \hfill
    \subfigure[Error propagation aware (EPA) training]{
        \includegraphics[width=0.4\linewidth]{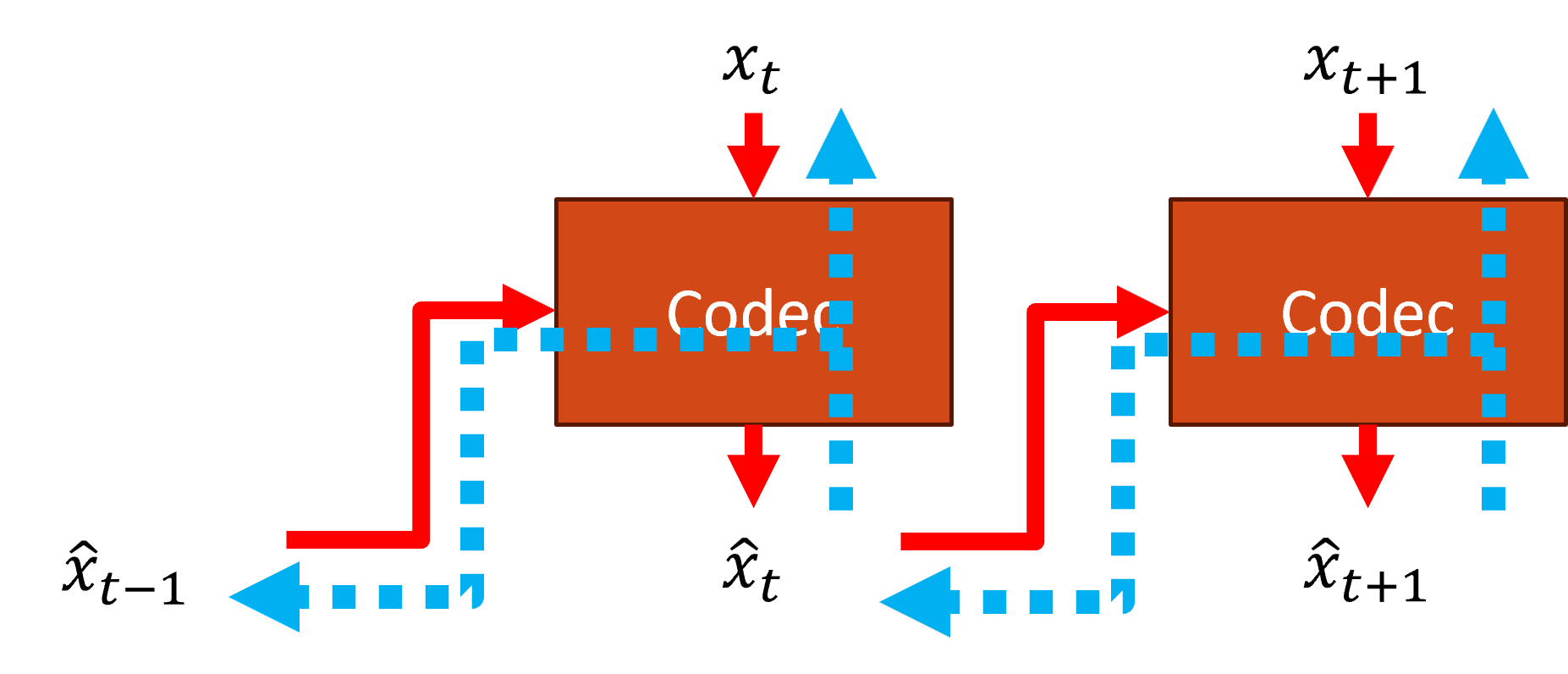}
        \label{fig:w_EPA}
    }
    \caption{Illustration of error propagation aware (EPA) training. Red anchor represents the forward path for encoding consecutive frames in training sequences. Blue anchor represents the gradient flows through the video codec.}
    \label{fig:EPA_compare}
\end{figure*}

As a first improvement, we incorporate the EPA training procedure into the baseline model. Inspired by Lu~\textit{et al.}~\cite{Lu2020ContentAA}, EPA training considers the dependencies among reconstructed P-frames in the training objective. To illustrate the difference, we compare the application of the EPA training procedure with not using it in Figure~\ref{fig:EPA_compare}. In the training of DVC~\cite{dvclu} or the baseline model by Ho~\textit{et al.}~\cite{canfvc}, each P-frame is updated separately. In contrast, EPA training updates the entire training sequences at once, taking into account the fact that the previously reconstructed frame serves as the reference frame for encoding the next P-frame.

By considering the dependencies among reconstructed frames, EPA training allows the codec to learn behavior that is closer to testing time. Introducing EPA training to the baseline model not only improves coding performance but also avoids introducing any extra model complexity.

\textbf{Improvement: Round-based Training}
Another improvement to achieve better alignment between training and inference time is round-based training. Conventional approaches, including our baseline model, follow Ballé~\textit{et al.}~\cite{googleiclr18} and adopt additive uniform noise to the latent variables to approximate Shannon cross entropy during training. However, during inference, the latents still need to be quantized for entropy coding.

To mitigate this mismatch, Minnen and Singh~\cite{minnen2020channelwise} propose round-based training for image compression. In round-based training, actual rounding operations are applied to the latents during the decoding transforms to reconstruct the image, while a straight-through estimator handles the gradient flow during backpropagation. By using additive uniform noise for entropy estimation only, round-based training can still learn a differential entropy model while better aligning the decoding process with inference time. This technique has been widely adopted in recent advancements in learned video compression~\cite{ssf, tcm, acmmm22, li2023neural}. We include round-based training in the baseline model to keep pace with these recent advances.

By incorporating EPA training and round-based training into the baseline model, we aim to achieve better alignment between the training and testing stages, leading to improved coding performance without introducing additional complexity. These advancements in training techniques enhance the overall performance and robustness of the video codec in real-world scenarios.

\subsection{Implementation Details}
To incorporate the various tools into the baseline framework, we follow a step-by-step process outlined below.

\textbf{Step 1: EPA Training and Round-based Training}
We first introduce EPA training and round-based training to the baseline model. This step involves fine-tuning the existing model using the pre-trained weights from Ho~\textit{et al.}~\cite{canfvc} while keeping the model architecture unchanged. We fine-tune the model using the highest rate model ($\lambda=2048$) with a learning rate of $10^{-5}$ and a batch size of 4. We perform EPA training and round-based training for 1 epoch and then fine-tune the model for lower rate models ($\lambda=1024, 512, 256$) for an additional 1 epoch.

\textbf{Step 2: Multi-scale Motion Compensation}
The next step involves incorporating the multi-scale motion compensation network~\ref{fig:gridnet} into the framework. We start with 2-frame (IP) training, following the protocol suggested in the supplementary material of Ho~\textit{et al.}~\cite{canfvc}. During this stage, we train the multi-scale motion compensation network while keeping other model weights fixed. The training objective is to minimize the distortion between the output of the multi-scale motion compensation network ($x_c$) and the coding frame, along with the rate of the motion codec. After the 2-frame training, we introduce the inter-frame codec and continue with the 2-frame training stage. Finally, we incorporate 5-frame (IPPPP) training to optimize the entire codec end-to-end. We gradually adjust the learning rate from $10^{-4}$ to $10^{-5}$ during this stage.

\textbf{Step 3: Modulated Loss with Feature Map Modulation}
In this step, we introduce additional modulation parameters into the framework to incorporate modulated loss. These parameters affect the intermediate feature maps of motion compensation and inter-frame coding. We follow a similar training process as in Step 2, starting with 2-frame and 5-frame training stages. Additionally, we include a 7-frame training stage with a batch size of 2 to fully utilize the training sequences.

\textbf{Step 4: Quadtree Partition-Based Entropy Model}
The final improvement we introduce is the quadtree partition-based entropy model. We directly apply the 7-frame training stage to train this model since the context model mainly influences the rate and not the latent itself. To ensure stability during training, we pre-train the quadtree partition-based entropy model in the conditional motion codec and conditional inter-frame codec by updating only their network weights. We train the model for 2 epochs to update the quadtree partition-based entropy model exclusively. Subsequently, we fine-tune the entire codec end-to-end for an additional 3 epochs, gradually adjusting the learning rate from $5\times 10^{-5}$ to $10^{-5}$. Finally, we fine-tune the resulting model for lower rate models ($\lambda=1024, 512, 256$) for another 1 epoch.

By following this step-wise implementation process, we seamlessly incorporate the proposed enhancements into the baseline framework, leading to improved performance and robustness in real-world video compression scenarios.
\section{Experiment}
\label{sec:experiment}
\subsection{Experimental Settings}
\subsubsection{Training Dataset and Other Details}
In training our model, we adhere to the common practice in end-to-end video compression research~\cite{dvclu, mlvc, dcvc, acmmm22, mimt} by utilizing the Vimeo-90k dataset~\cite{vimeo}. The input sequences are randomly cropped and flipped to a resolution of $256 \times 256$, and all $7$ frames are utilized for training. We adopt the Adam optimizer~\cite{Adam} and train using two NVIDIA Tesla V100 GPUs. The training framework is implemented using PyTorch 1.4.0~\cite{pytorch140}.

\subsubsection{Test Conditions}
To comprehensively evaluate the performance of our proposed methods, we conducted extensive experiments on well-established benchmark datasets, including UVG~\cite{uvg}, HEVC Class B~\cite{hevcctc}, and MCL-JCV~\cite{mcl}. 
To showcase the inter-frame performance, we adopted a long Group-of-Pictures (GOP) setting of 32, following recent studies~\cite{tcm, canfvc, acmmm22, mimt}. The testing sequences were converted from YUV420 format to RGB format using FFmpeg~\cite{ffmpeg}. Each testing sequence consists of a total of $96$ frames.

\begin{table*}[t]
    \caption{Incremental BD-Rate reduction achieved by methods integrated into the baseline framework. Each column represents a setting that includes the method with a "check" mark into the baseline framework. Our proposed CANF-VC++ is the model of Setting (d). \textbf{Anchor}: Ho~\textit{et al.}~\cite{canfvc}.}
    \centering
    \small
    \begin{tabular}{c|c|cccc}
        \toprule
        \multicolumn{2}{c|}{Methods} &(a)&(b)&(c)&(d) \\
        \hline
        \multicolumn{2}{c|}{EPA \& Round-based Training}                &\checkmark & \checkmark & \checkmark & \checkmark \\
        \multicolumn{2}{c|}{Multi-scale MCNet}                          &           & \checkmark & \checkmark & \checkmark \\
        \multicolumn{2}{c|}{Modulated Loss with Feature Map Modulation} &           &            & \checkmark & \checkmark \\
        \multicolumn{2}{c|}{Quadtree Partition-Based Entropy Model}     &           &            &            & \checkmark \\
        \hline\hline
        \multirow{2}{*}{} BD-rate (\%) & \multirow{1}{*}{UVG}           & -8.2      & -17.0      & -25.6      & -40.2\\
             & \multirow{1}{*}{HEVC-B}                                  & -8.7      & -8.4       & -23.8      & -38.1\\
             & \multirow{1}{*}{MCL-JCV}                                 & -9.1      & -13.9      & -22.0      & -35.5\\
        \bottomrule
    \end{tabular}
    \label{tab:increamental_gain}
\end{table*}

The rate-distortion performance is assessed using two key metrics: the bit-rate (bits-per-pixel, bpp) and the Peak Signal-to-Noise Ratio in RGB domain (PSNR-RGB). The PSNR-RGB is calculated using the formula:
\begin{equation}
\text{PSNR}_{\text{RGB}} = 10 \cdot \log{10}\left(\frac{{\text{MAX}^2}}{{\text{MSE}_{\text{RGB}}}}\right)
\end{equation}
where $\text{PSNR}_{\text{RGB}}$ represents the PSNR value in the RGB domain. $\text{MAX}$ is the maximum possible pixel value (e.g., 255 for 8-bit color depth), and $\text{MSE}_{\text{RGB}}$ is the mean squared error between the original RGB frame and the reconstructed frame.

For further evaluation, we calculate the BD-rate~\cite{bdrate} of each method with respect to a reference point known as the "Anchor." The BD-rate allows us to measure the coding efficiency improvements achieved by our proposed methods compared to the reference point, considering both bit-rate reduction and PSNR-RGB gains. The average bpp and PSNR-RGB are computed across the entire dataset, treating it as a single input sequence.

\subsection{Rate-Distortion Performance}
\label{exp:rd}
\subsubsection{Incremental Enhancements of Additional Tools}
The rate-distortion performance of the incremental enhancements integrated into the baseline framework by Ho~\textit{et al.}~\cite{canfvc} is presented in Table~\ref{tab:increamental_gain}. The table showcases the cumulative BD-rate reduction achieved by incorporating multiple methods compared to the baseline framework. The methods include EPA training and round-based training, multi-scale motion compensation, modulated loss with feature map modulation, and quadtree partition-based entropy model. When all methods are integrated, a substantial improvement in BD-rate reduction is observed across all datasets. The cumulative BD-rate reduction reaches -40.2\% on the UVG dataset, -38.1\% on the HEVC Class B dataset, and -35.5\% on the MCL-JCV dataset. These results demonstrate the synergistic effect of integrating multiple methods, which collectively enhance the rate-distortion performance of the baseline framework.

Of all the methods, the "Quadtree Partition-Based Entropy Model" stands out as the most impactful in terms of performance improvements, whereas the "Multi-scale MCNet" shows the least significant enhancement. This observation underscores the significant potential for improving entropy coding efficiency within the baseline framework CANF-VC~\cite{canfvc}, while the exploration of feature domain conditioning signal formulation appears to be comparatively less crucial. The remaining two tools, "EPA \& Round-based Training" and "Modulated Loss with Feature Map Modulation," contribute notable performance gains while incurring minimal additional model complexity, as detailed in Section~\ref{ssec:complexity}. In summary, the integration of each of these tools has effectively elevated the performance of the baseline frameworks, thereby advancing coding efficiency.

\begin{figure*}
  \centering
  \subfigure[UVG~\cite{uvg}]{
    \includegraphics[width=0.3\linewidth]{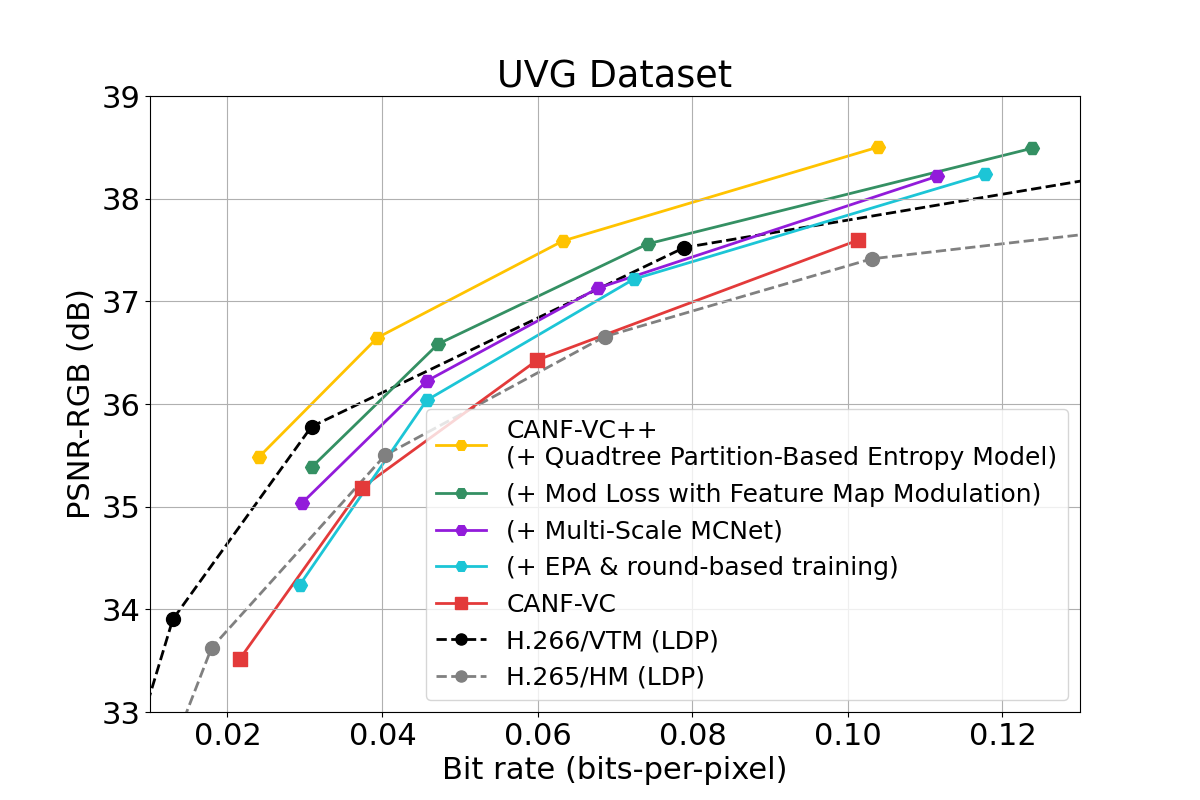}
    \label{fig:rd_uvg}
  }
  \hfill
  \subfigure[HEVC Class B~\cite{hevcctc}]{
    \includegraphics[width=0.3\linewidth]{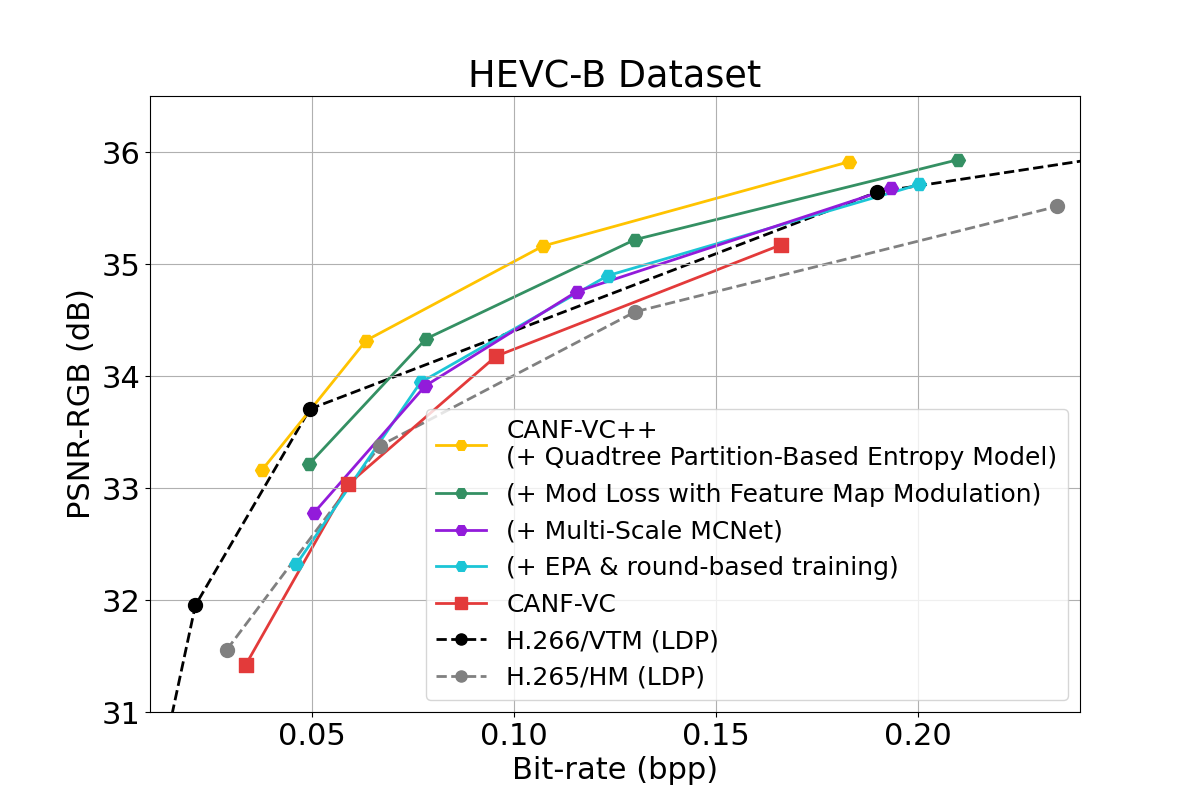}
    \label{fig:rd_hevcb}
  }
  \hfill
  \subfigure[MCL-JCV~\cite{mcl}]{
    \includegraphics[width=0.3\linewidth]{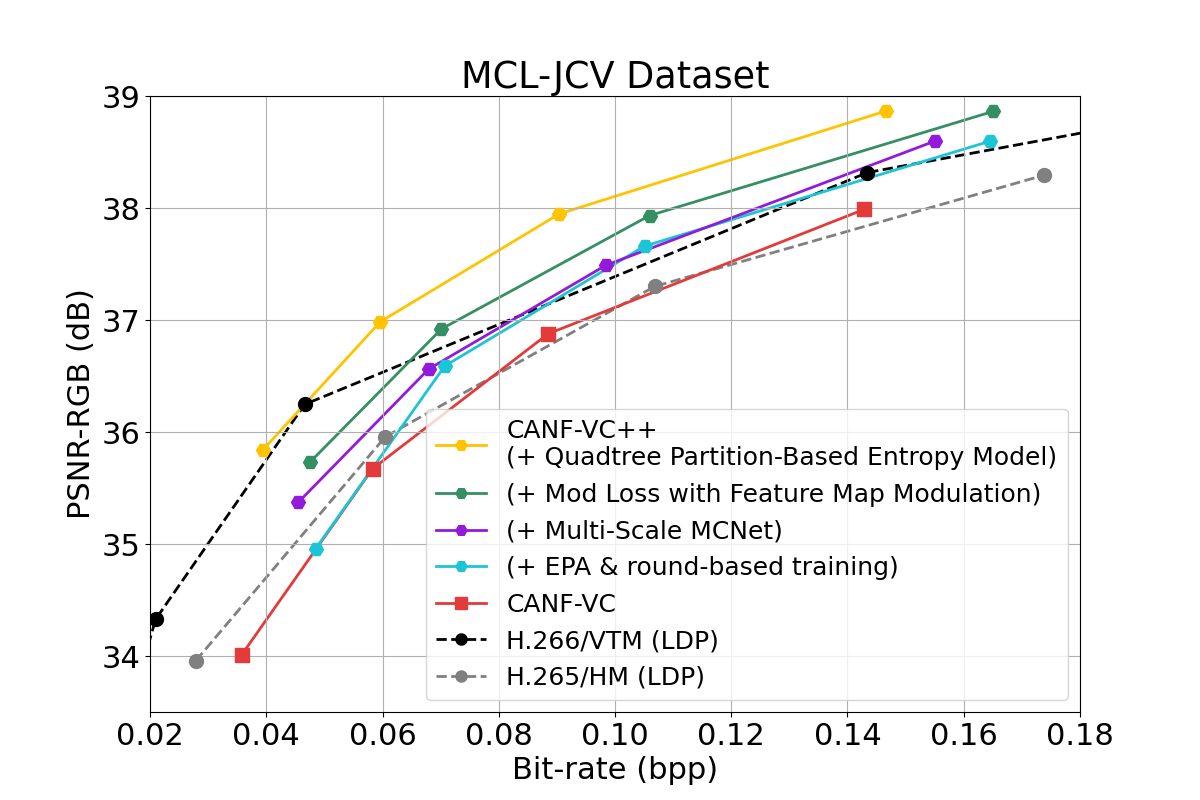}
    \label{fig:rd_mcl}
  }
  \caption{Rate-Distortion performance comparison of methods applied incrementally to baseline framework~\cite{canfvc} on different datasets.}
  \label{fig:rd}
\end{figure*}

Figure~\ref{fig:rd} illustrates the rate-distortion (RD) curves of the methods presented in Table~\ref{tab:increamental_gain}, providing a visual representation of their performance improvement. The cyan curve represents the performance with EPA training and round-based training, which significantly improves the overall coding performance, except for the lowest rate point. The purple curve represents the performance with multi-scale motion compensation network, which enhances coding performance, particularly at low rates. The green curve represents the performance with modulated loss and feature map modulation, surpassing the performance of the H.266/VVC reference software (VTM)~\cite{VVCSoftware_VTM} at high rates. The introduction of the quadtree partition-based entropy model further reduces the bit-rate across all working points, resulting in improved rate-distortion performance.

\subsubsection{Compare with Other Learned Video Compression Methods}
We also compare our CANF-VC++ with recent publications, including DCVC~\cite{dcvc} (NIPS'21), DCVC-TCM~\cite{tcm} (TMM'22), DCVC-HEM~\cite{acmmm22} (ACM MultiMedia'22), and DCVC-DC~\cite{li2023neural} (CVPR'23). We adopt GOP of 32, testing sequence length of 96 for fair comparison. We use their released testing software to run the evaluation. As shown in Table~\ref{tab:bd_rate_sota} and Figure~\ref{fig:rd_sota}, Our CANF-VC++ is better than DCVC~\cite{dcvc} and DCVC-TCM~\cite{tcm} but still worse than DCVC-TCM~\cite{tcm} and DCVC-DC~\cite{li2023neural}. It is worse noting that our system, which extends from the baseline framework Ho~\textit{et al.}~\cite{canfvc}, provides an alternative learned video codec with far smaller buffer size than DCVC-TCM~\cite{tcm}, DCVC-TCM~\cite{tcm}, and DCVC-DC~\cite{li2023neural}, as shown in Table~\ref{tab:complexity}. We expect our system can be further improved by carefully combing techniques from recent advances.

\begin{table}
    \caption{BD-Rate comparison of Ho~\textit{et al.}~\cite{canfvc} (CANF-VC) and our proposed CANF-VC++ with other learned video compression works, including DCVC~\cite{dcvc}, DCVC-TCM~\cite{tcm}, DCVC-HEM~\cite{acmmm22}, and DCVC-DC~\cite{li2023neural}. \textbf{Anchor}: Ho~\textit{et al.}~\cite{canfvc} (CANF-VC).}
    \centering
    \small
    \begin{tabular}{c|ccc}
        \toprule
        \multirow{2}{*}{Methods}                   & \multicolumn{3}{c}{BD-rate (\%)} \\
        \cline{2-4}
                                                   &  UVG           &  HEVC-B    & MCL-JCV                        \\
        \hline
        CANF-VC~\cite{canfvc} (anchor)             &  -             & -          & -        \\
        \textbf{CANF-VC++} (Ours)                  & -40.2          & -38.1      & -35.5    \\
        DCVC~\cite{dcvc}                           & 29.6           & 21.4       & 13.6     \\
        DCVC-TCM~\cite{tcm}                        & -22.8          & -23.8      & -20.9    \\
        DCVC-HEM~\cite{acmmm22}                    & -44.5          & -42.8      & -41.6    \\
        DCVC-DC~\cite{li2023neural}                 & -52.2          & -50.0      & -49.7    \\
        
        \bottomrule
    \end{tabular}
    \label{tab:bd_rate_sota}
\end{table}

\begin{figure*}
  \centering
  \subfigure[UVG~\cite{uvg}]{
    \includegraphics[width=0.3\linewidth]{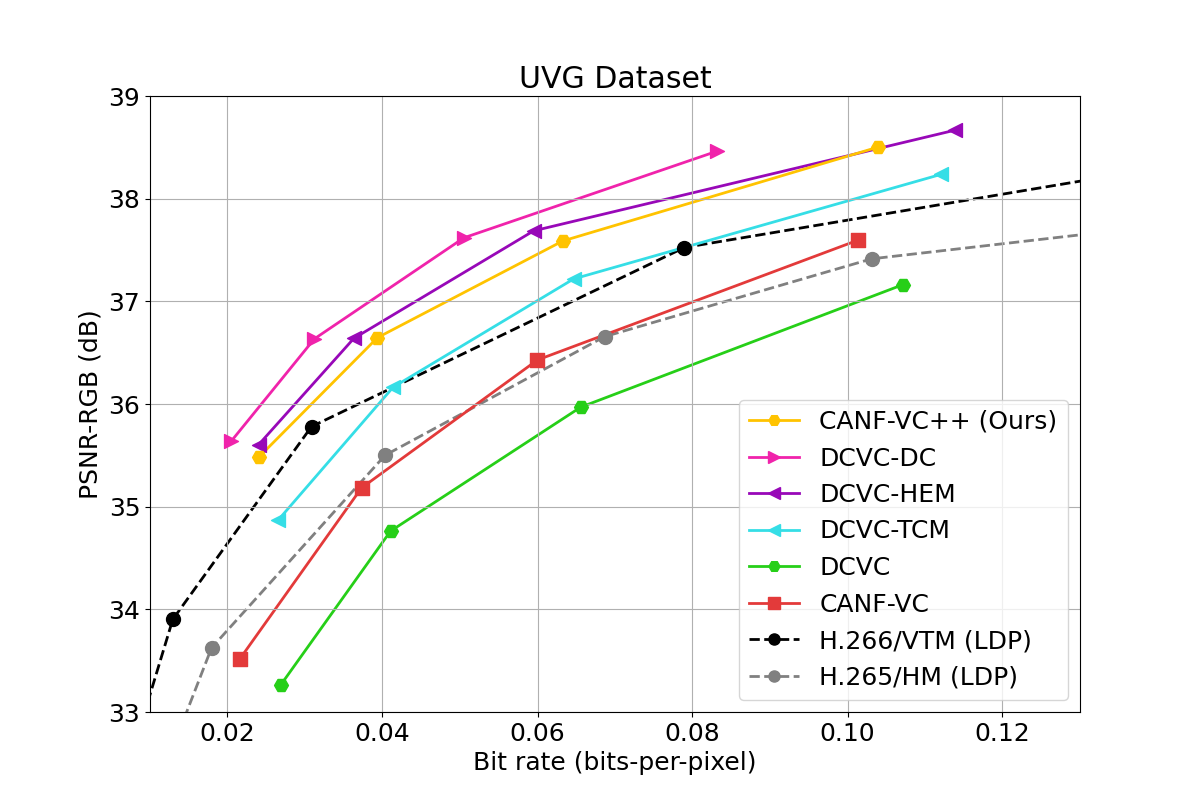}
    \label{fig:rd_sota_uvg}
  }
  \hfill
  \subfigure[HEVC Class B~\cite{hevcctc}]{
    \includegraphics[width=0.3\linewidth]{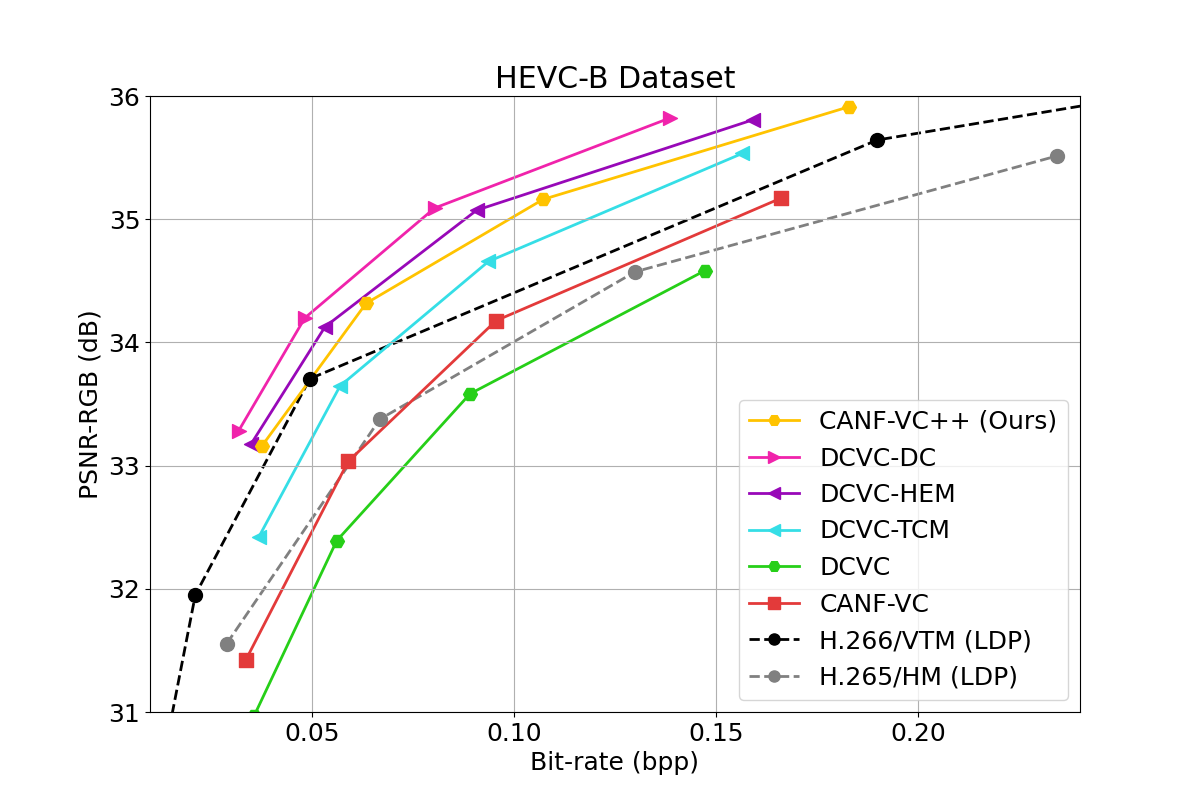}
    \label{fig:rd_sota_hevcb}
  }
  \hfill
  \subfigure[MCL-JCV~\cite{mcl}]{
    \includegraphics[width=0.3\linewidth]{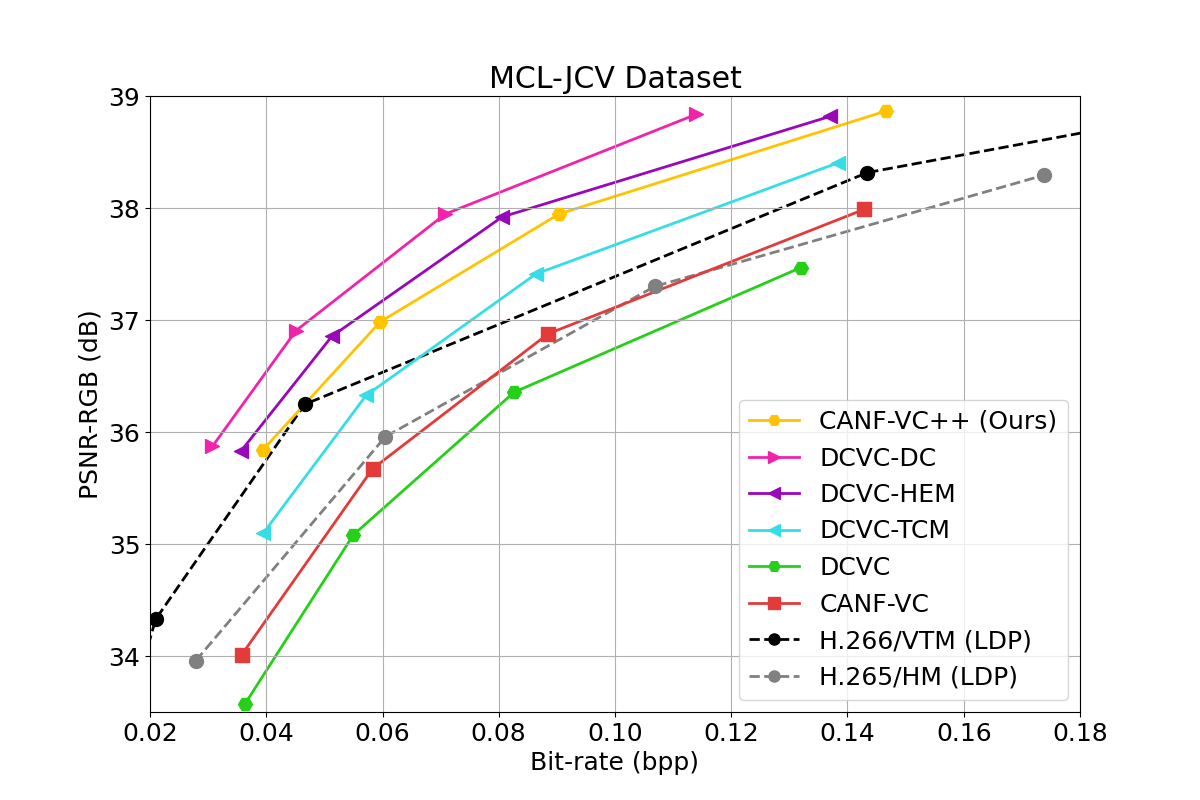}
    \label{fig:rd_sota_mcl}
  }
  \hfill
  \caption{Rate-Distortion performance comparison of methods, including baseline framework~\cite{canfvc} (CANF-VC), our proposed CANF-VC++, DCVC~\cite{dcvc}, DCVC-TCM~\cite{tcm}, DCVC-HEM~\cite{acmmm22}, and DCVC-DC~\cite{li2023neural} on different datasets.}
  \label{fig:rd_sota}
\end{figure*}

\subsubsection{Compare with H.265~\cite{HM}/H.266~\cite{VVCSoftware_VTM}}
Additionally, we include the results from testing software of traditional codecs H.265~\cite{HM} and H.266~\cite{VVCSoftware_VTM} under the low-delay P (LDP) configuration for comparison in Figure~\ref{fig:rd} and Table~\ref{tab:bd_rate_vtm}. Comparing to the baseline model that performs close to H.265/HM, our improved model can outperform H.266 by a significant margin, showcasing the success of our integration of tools into the baseline model.

These experimental results highlight the effectiveness of the incremental enhancements and their collective impact on the rate-distortion performance of the video codec. The introduced methods exhibit superior performance compared to traditional codecs, demonstrating their potential for practical video compression scenarios.

\begin{table}
    \caption{BD-Rate comparison of Ho~\textit{et al.}~\cite{canfvc} (CANF-VC) and our proposed CANF-VC++ with H.266/VTM~\cite{VVCSoftware_VTM} and H.265/HM~\cite{HM} under LDP configuration. \textbf{Anchor}: VTM 20.0~\cite{VVCSoftware_VTM}}
    \centering
    \small
    \begin{tabular}{c|ccc}
        \toprule
        \multirow{2}{*}{Methods}                   & \multicolumn{3}{c}{BD-rate (\%)} \\
        \cline{2-4}
                                                   &  UVG           &  HEVC-B    & MCL-JCV                        \\
        \hline
        VTM 20.0~\cite{VVCSoftware_VTM} (anchor)   &  -             & -          & -        \\
        HM 16.22~\cite{HM}                         & 49.4           & 53.9       & 42.2     \\
        \hline
        CANF-VC~\cite{canfvc}                      & 55.4           & 59.6       & 53.1     \\
        \textbf{CANF-VC++} (Ours)                  & -20.1          & -14.7      & -15.3    \\
        
        \bottomrule
    \end{tabular}
    \label{tab:bd_rate_vtm}
\end{table}

\subsection{Ablation Study}
In this section, we conduct an ablation study to examine the effects of modulated loss and feature map modulation on the compression performance.

\subsubsection{Modulated Loss}
\label{exp:ablation_mod_loss}
We begin by analyzing the impact of modulated loss and feature map modulation. Starting from the model that includes EPA training, round-based training, and multi-scale MCNet in the baseline framework, we compare different schemes: applying modulated loss only, applying feature map modulation only, and using both modulated loss and feature map modulation simultaneously. Table~\ref{tab:mod_loss} presents the BD-rate of these schemes. It is evident that applying modulated loss or feature map modulation alone does not lead to a significant improvement in performance. However, when feature map modulation is incorporated, we observe substantial performance gains, indicating that feature map modulation unlocks the potential of modulated loss. It is worth noting that in our comparison with Table~\ref{tab:modules_to_be_adapted}, feature map modulation is applied only to the multi-scale MCNet, while the conditional inter-frame codec remains un-modulated.

\begin{table}
  \caption{Impact of modulated loss and feature map modulation on compression performance. \textbf{Anchor}: Setting (c) in Table~\ref{tab:increamental_gain}, which is baseline model with EPA training, round-based training, and multi-scale MCNet but without feature map modulation and modulated loss.}
  \begin{center}{
      \centering
        \begin{tabular}{c|c|ccc}
            \toprule
            \multicolumn{2}{c|}{Feature Map Modulation}              &            & \checkmark & \checkmark\\
            \multicolumn{2}{c|}{Modulated Loss}                      & \checkmark &            & \checkmark\\
            \cline{2-2}
            \hline\hline
            \multirow{2}{*}{} BD-rate (\%) & \multirow{1}{*}{UVG}    & 1.1        & -2.9       & -10.9     \\
                                           & \multirow{1}{*}{HEVC-B} & 3.0        & -2.2       & -12.6     \\
                                           & \multirow{1}{*}{MCL-JCV}& -0.7       & -2.5       & -9.0      \\
            \bottomrule
        \end{tabular}
  }\end{center}
  \label{tab:mod_loss}
\end{table}

\subsubsection{Effect of Error-Propagation-Aware (EPA) Training and Modulated Loss}
\label{exp:ablation_epa_mod_loss}

\begin{figure*}
  \centering
  \subfigure[UVG~\cite{uvg}]{
    \includegraphics[width=0.3\linewidth]{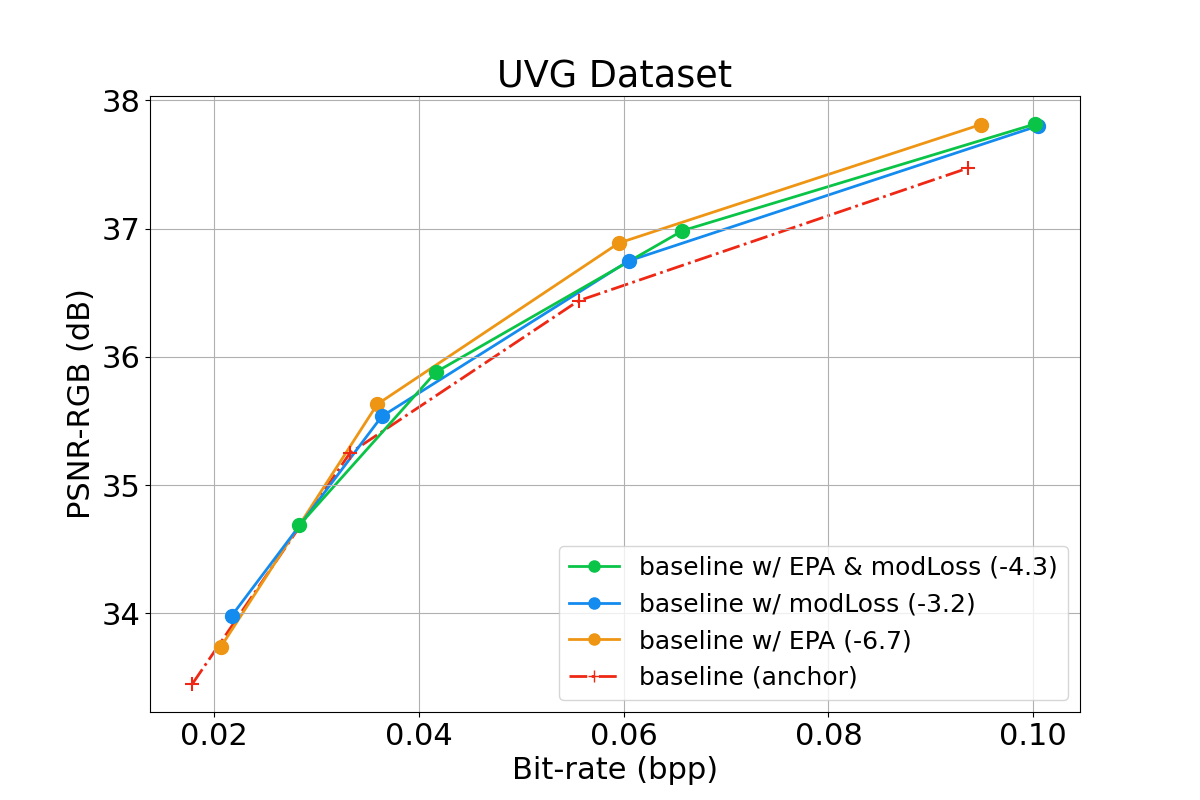}
    \label{fig:rd_EPA_modLoss_uvg}
  }
  \hfill
  \subfigure[HEVC Class B~\cite{hevcctc}]{
    \includegraphics[width=0.3\linewidth]{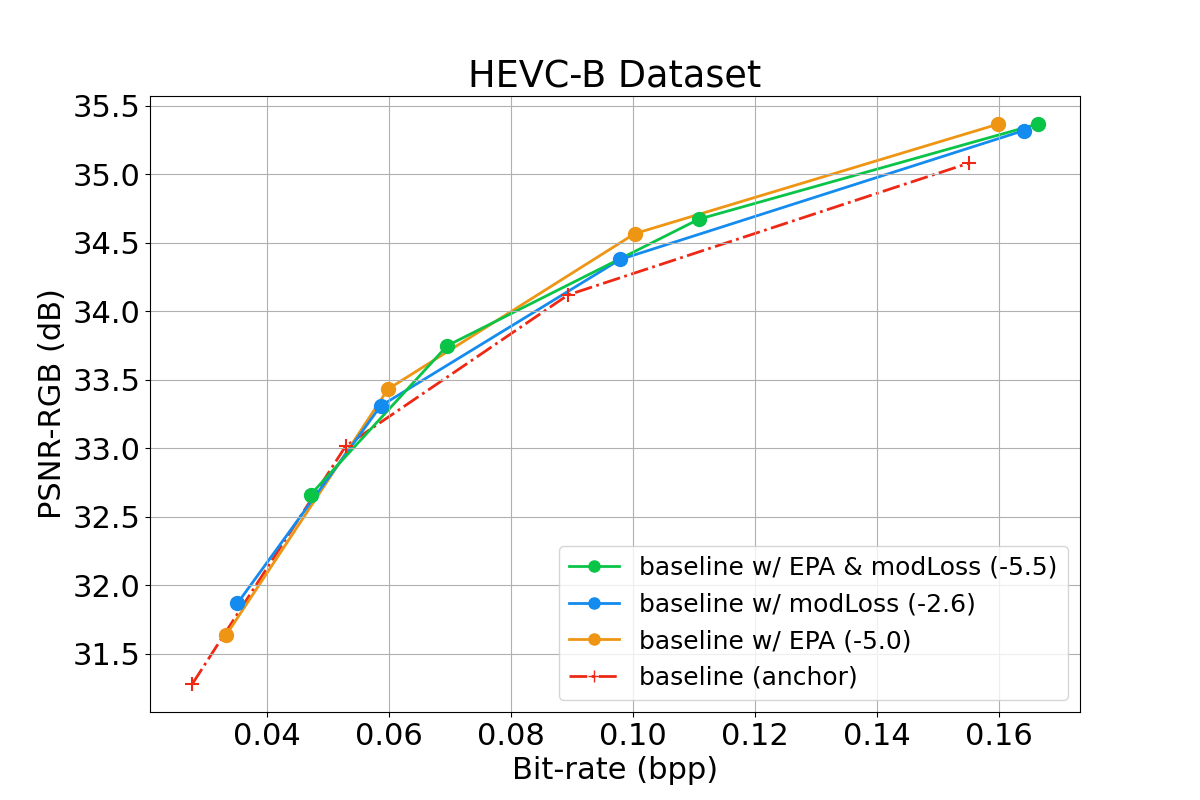}
    \label{fig:rd_EPA_modLoss_hevcb}
  }
  \hfill
  \subfigure[MCL-JCV~\cite{mcl}]{
    \includegraphics[width=0.3\linewidth]{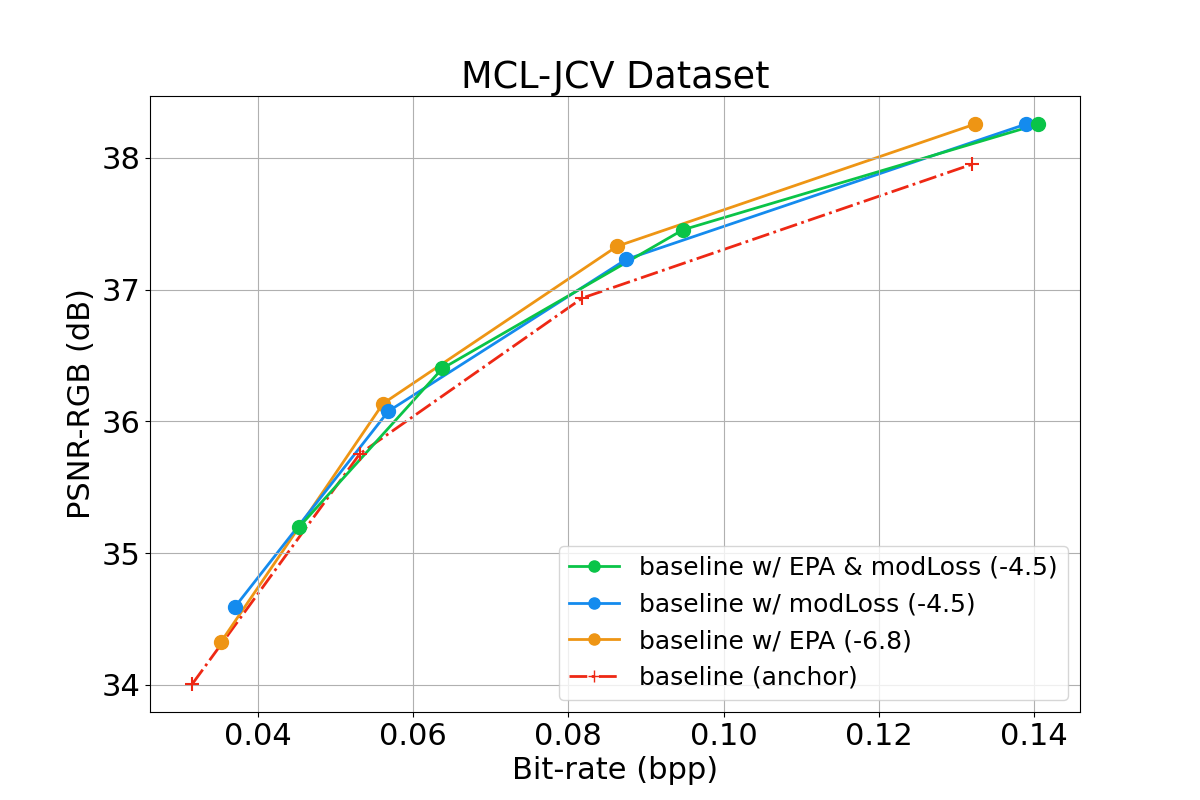}
    \label{fig:rd_EPA_modLoss_mcl}
  }
  \caption{Comparison of the rate-distortion performance when employing the Error-Propagation-Aware (EPA) training strategy, the Modulated Loss technique, and both of them in combination. BD-rate (\%) compared to the anchor method is indicated in parentheses. \textbf{Anchor}: Ho~\textit{et al.}~\cite{canfvc} (CANF-VC) with round-based training.}
  \label{fig:rd_EPA_modLoss}
\end{figure*}

In this study, we conduct a comprehensive comparison of rate-distortion performance while employing Error-Propagation-Aware (EPA) training, the Modulated Loss technique. Both techniques are designed to mitigate error-propagation issues in learned P-frame coding schemes. Our objective is to assess whether these two training methods can synergize to enhance performance or if they counteract each other's benefits.

We apply EPA training and Modulated Loss individually to the baseline framework. Figure~\ref{fig:rd_EPA_modLoss} compares rate-distortion performance of applying EPA training strategy only, applying modulated loss only, and applying both to baseline framework. The BD-rate (\%) of each method to baseline framework as anchor is shown in parentheses. Our observations reveal that applying either EPA training or Modulated Loss yields performance improvements, demonstrating the effectiveness of these techniques. However, no further enhancements are observed when both EPA training and Modulated Loss are concurrently applied.A similar conclusion can be drawn in Section~\ref{exp:ablation_mod_loss}, where we demonstrate that solely applying Modulated Loss does not yield improvements when the anchor method already includes EPA training.

\begin{figure*}
  \centering
  \subfigure[Rate-distortion performance comparison on sequence \textit{Cactus}.]{
    \includegraphics[width=0.55\linewidth]{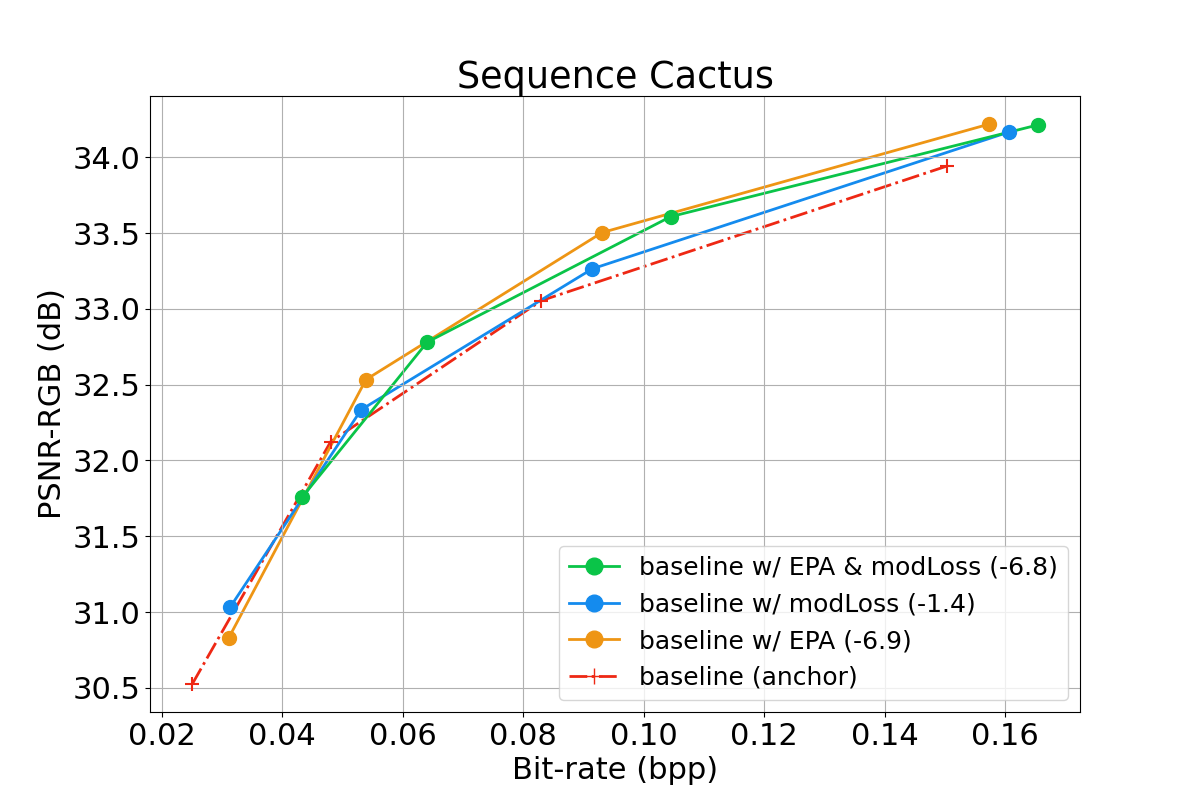}
    \label{fig:rd_Cactus}
  }
  \hfill
  \subfigure[PSNR and bit-rate profile at high bit-rate ($\lambda=2048$). I-frames omitted for improved visibility of P-frame performance.]{
    \includegraphics[width=0.6\linewidth]{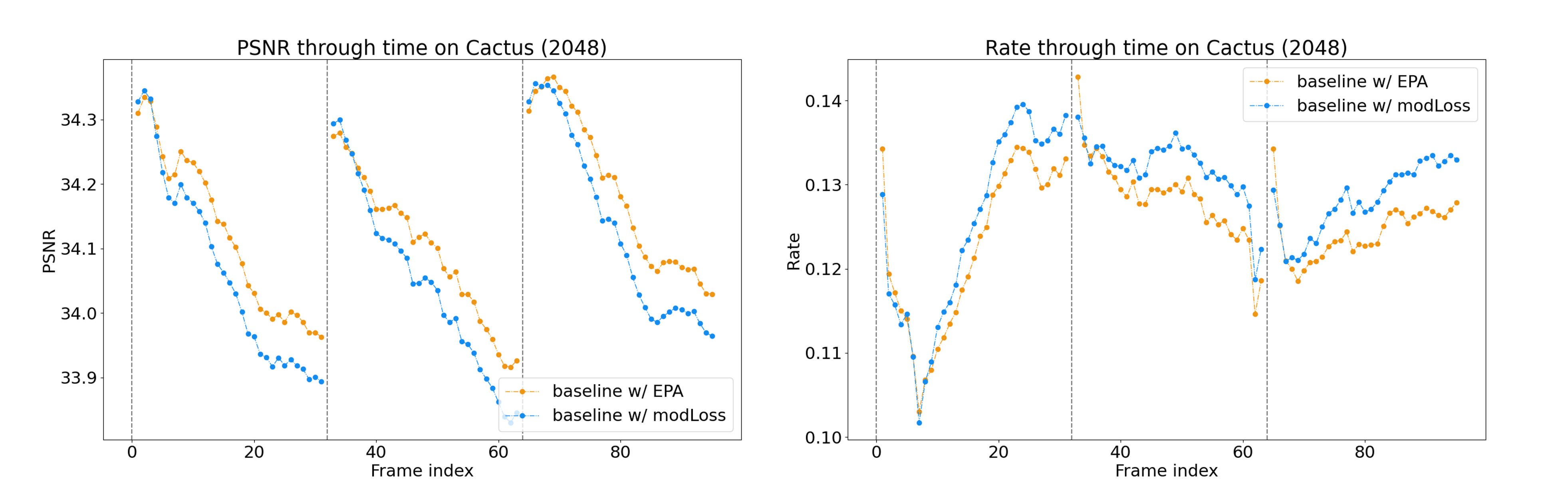}
    \label{fig:profile_Cactus_high}
  }
  \hfill
  \subfigure[PSNR and bit-rate profile at low bit-rate ($\lambda=256$). I-frames omitted for improved visibility of P-frame performance.]{
    \includegraphics[width=0.6\linewidth]{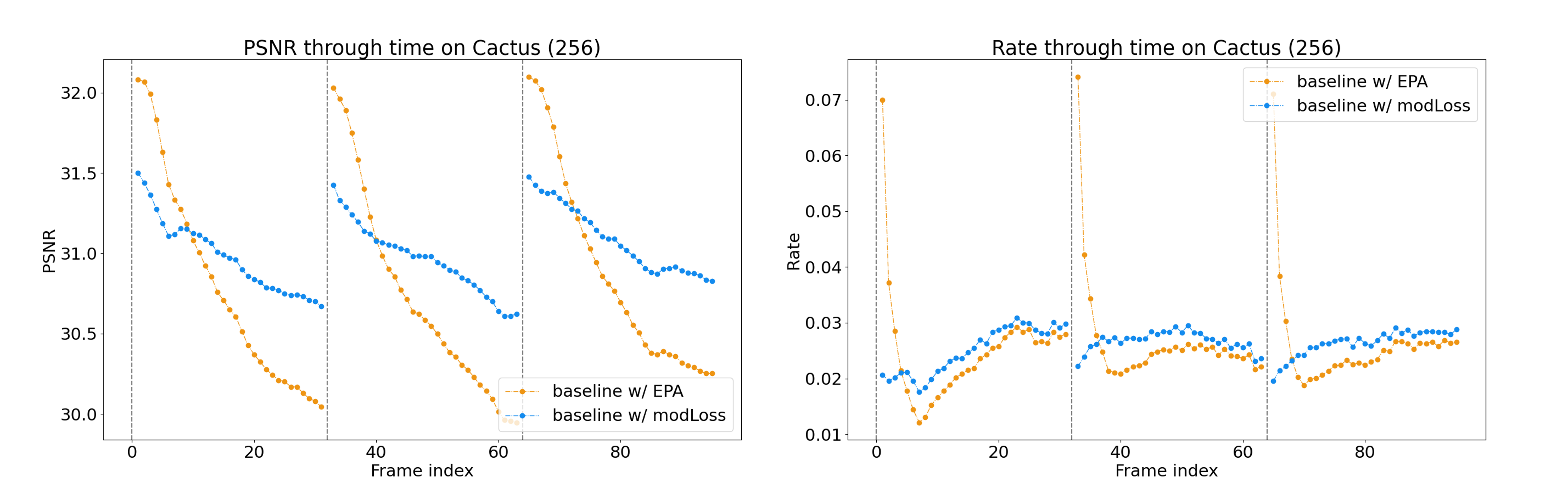}
    \label{fig:profile_Cactus_low}
  }
  \caption{Sequence-level rate-distortion performance along with per-frame PSNR and bit-rate profile on sequence~\textit{Cactus}.}
  \label{fig:rd_profile_Cactus}
\end{figure*}

To gain deeper insights, we investigate sequence-level rate-distortion curves, per-frame PSNR profiles, and bit-rate profiles, as shown in Figure~\ref{fig:rd_profile_Cactus}.  Specifically, in Figure~\ref{fig:rd_Cactus}, we analyze the rate-distortion performance on the \textit{Cactus} sequence. In Figure~\ref{fig:profile_Cactus_high} (high bit-rate, $\lambda=2048$), and Figure~\ref{fig:profile_Cactus_low} (low bit-rate, $\lambda=256$), we examine the per-frame PSNR and bit-rate profiles.

Notably, we observe that the baseline with EPA training tends to allocate more bits to the initial frames, while the baseline with Modulated Loss allocates more bits to later frames. We posit that EPA training, by back-propagating through all frames during training, benefits from allocating more bits to the early frames, which, in turn, improves overall coding performance. Conversely, Modulated Loss emphasizes the distortion term's importance for coding later frames, leading to a preference for allocating more bits to them. These strategies are counteractive, resulting in sub-optimal performance when combined. Furthermore, we notice that EPA training is more effective at high bit-rates, while Modulated Loss performs better at low bit-rates. The comparison in Figure~\ref{fig:profile_Cactus_high} and Figure~\ref{fig:profile_Cactus_low} demonstrates that EPA training benefits high bit-rate coding ($\lambda=2048$) but is less efficient at low bit-rates ($\lambda=256$), where error-propagation effects are more pronounced. In contrast, Modulated Loss exhibits less PSNR degradation and superior coding performance by allocating more bits to later frames.

In conclusion, EPA training and Modulated Loss represent two distinct training strategies for mitigating error-propagation. They differ in their bit-allocation strategies for Group-of-Pictures (GOP). Simply applying both simultaneously to the learned P-frame codec training procedure does not yield improved overall coding performance. This suggests the need for further investigation into optimizing bit-allocation for different rate ranges and video contents. Notably, our proposed feature map modulation with Modulated Loss (Section~\ref{ssec:drift_error} and Section~\ref{exp:ablation_mod_loss}) presents a more effective way to harness the potential synergy between these two training techniques.

\subsubsection{Feature Map Modulation on Different Submodules}
\label{exp:modules_to_be_adapted}
Next, we investigate the impact of feature map modulation on different modules within the framework. Specifically, we examine the multi-scale feature extractor, GridNet, and the inter-frame codec. In our analysis, we divide the multi-scale motion compensation network, as shown in Figure~\ref{fig:gridnet}, into sub-modules of the multi-scale feature extractor and GridNet, and apply feature map modulation separately to each submodule. This approach allows us to assess the individual contributions of feature map modulation on these modules.

Table~\ref{tab:modules_to_be_adapted} presents the BD-rate results for the different modules when feature map modulation is applied. All methods are trained with the assistance of modulated loss. We observe that feature map modulation has varying effects on different modules. The multi-scale feature extractor exhibits a moderate improvement in performance, indicating that feature map modulation enhances the extraction of multi-scale features, contributing to better compression results. In contrast, GridNet demonstrates a more substantial enhancement, highlighting the significance of feature map modulation in improving spatial prediction accuracy. Furthermore, the inter-frame codec also benefits from feature map modulation, resulting in improved rate-distortion performance across all datasets.

In conclusion, our ablation study provides insights into the significance of modulated loss and feature map modulation in improving compression performance. By incorporating these techniques, particularly when applied together across different modules, we achieve notable enhancements in the rate-distortion performance of the video codec. The results highlight the importance of feature map modulation and its potential to improve various components within the baseline framework.

\begin{table}[ht]
    \caption{Impact of feature map modulation on different modules in the baseline framework. All of the methods are trained with the help of modulated loss. \textbf{Anchor}: Setting (c) in Table~\ref{tab:increamental_gain}, which is baseline model with EPA training, round-based training and multi-scale MCNet.}
    \begin{center}{
        \begin{tabular}{c|c|ccccc}
            \toprule
            \multicolumn{2}{c|}{Modules to be Adapted} & \multicolumn{4}{c}{} \\
            \hline
            \multicolumn{2}{c|}{Multi-scale Feature Extractor}   & \checkmark & \checkmark & \checkmark &           \\
            \multicolumn{2}{c|}{GridNet}                         &            & \checkmark & \checkmark &           \\
            \multicolumn{2}{c|}{Inter-frame Codec}               &            &            & \checkmark &           \\
            \multicolumn{2}{c|}{GridNet (last layer only)}       &            &            &            & \checkmark\\
            \hline\hline
            \multirow{2}{*}{} BD-rate (\%)& \multirow{1}{*}{UVG} & -2.6       & -10.9      & -9.8       &  -6.9     \\
                 & \multirow{1}{*}{HEVC-B}                       & -0.8       & -12.6      & -15.3      &  -9.8     \\
                 & \multirow{1}{*}{MCL-JCV}                      & -3.3       & -9.0       & -9.3       &  -6.8     \\
            \bottomrule
        \end{tabular}
    }\end{center}
    \label{tab:modules_to_be_adapted}
\end{table}

\subsubsection{Impact of Input for Feature Map Modulation}
\label{exp:ablation_state_signal}
In this ablation study, we aim to examine the impact of input for feature map modulation in our proposed method. Specifically, we investigate how changing the input from temporal propagated optical flow maps to manually-set signals affects the coding performance. This experiment allows us to understand the extent to which the choice of input influences the effectiveness of feature map modulation.

\begin{table}
    \caption{Impact of input for feature map modulation.The anchor method for BD-rate calculation is Setting (c) in Table~\ref{tab:increamental_gain}, which is baseline model with EPA training, round-based training and multi-scale MCNet. We test Setting (c) in Table~\ref{tab:increamental_gain} with different schemes to observe the influence of input for feature map modulation. \textbf{Anchor}: Setting (c) in Table~\ref{tab:increamental_gain}.}
    \centering
    \small
    \begin{tabular}{c|cc}
        \toprule
        \multirow{2}{*}{Methods}                   & \multicolumn{2}{c}{BD-rate (\%)} \\
        \cline{2-3}
                                                   &  UVG           &  HEVC-B                            \\
        \hline
        Reverse Order                              &  1.2           & 0.6                  \\
        Mis-match (\textit{BasketBallDrive})       &  -1.2          & -0.9                 \\
        Mis-match (\textit{HoneyBee})              &  -1.0          & -0.7                 \\
        
        \bottomrule
    \end{tabular}
    \label{tab:ablation_state_signal}
\end{table}

To evaluate this, we conduct two sets of experiments as outlined in Table~\ref{tab:ablation_state_signal}. In the first set, we reverse the order of temporal propagated optical flow maps in coding a Group-of-Pictures (GOP) sequence, denoted as "Reverse Order" in the table. We encode each testing sequence twice, once to record the temporal propagated optical flow maps and another time using the reverse order of these maps. In the second set, we use the temporal propagated optical flow maps from a fast motion sequence, \textit{BasketBallDrive}, for all the other sequences, labeled as "Mis-match (\textit{BasketBallDrive})". Similarly, we apply the temporal propagated optical flow maps from the sequence \textit{HoneyBee} with a static scene to all the other sequences, labeled as "Mis-match (\textit{HoneyBee})".

\begin{figure*}
  \centering
  \subfigure[Comparison of the maximum, mean, and minimum values altogether.]{
    \includegraphics[width=0.4\linewidth]{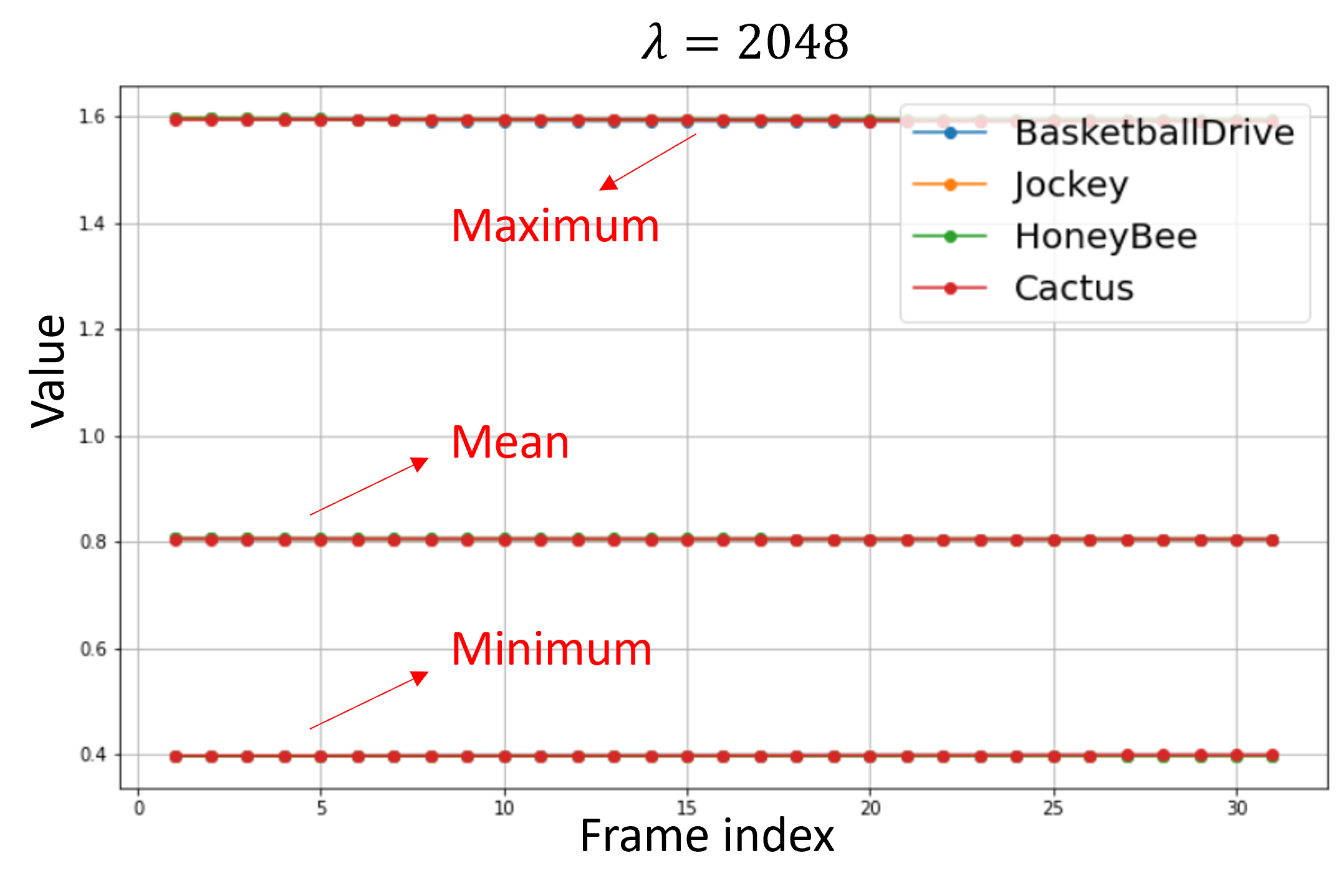}
  }
  
  \subfigure[Comparison of the maximum values.]{
    \includegraphics[width=0.4\linewidth]{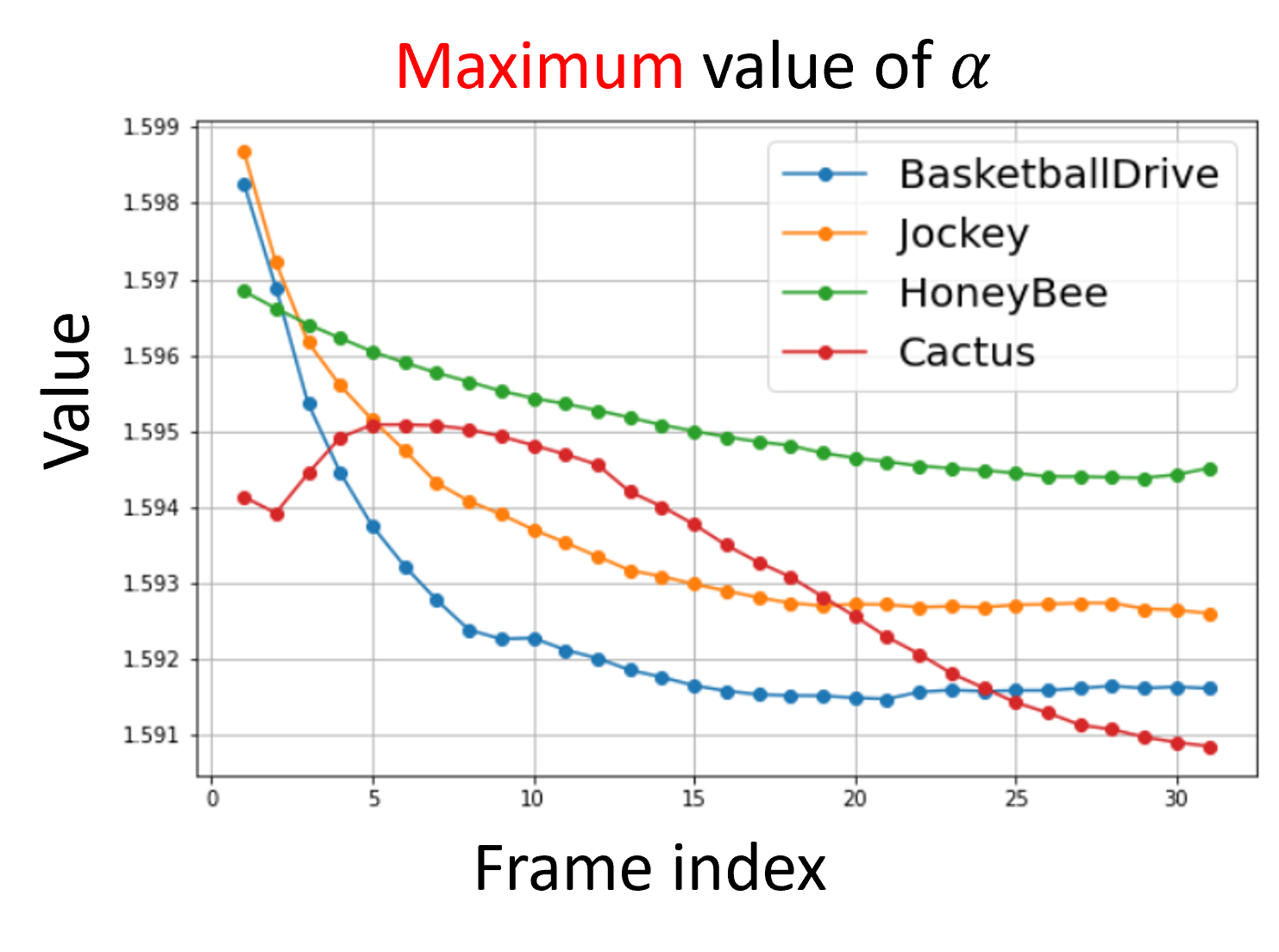}
  }
  
  \subfigure[Comparison of the mean values.]{
    \includegraphics[width=0.4\linewidth]{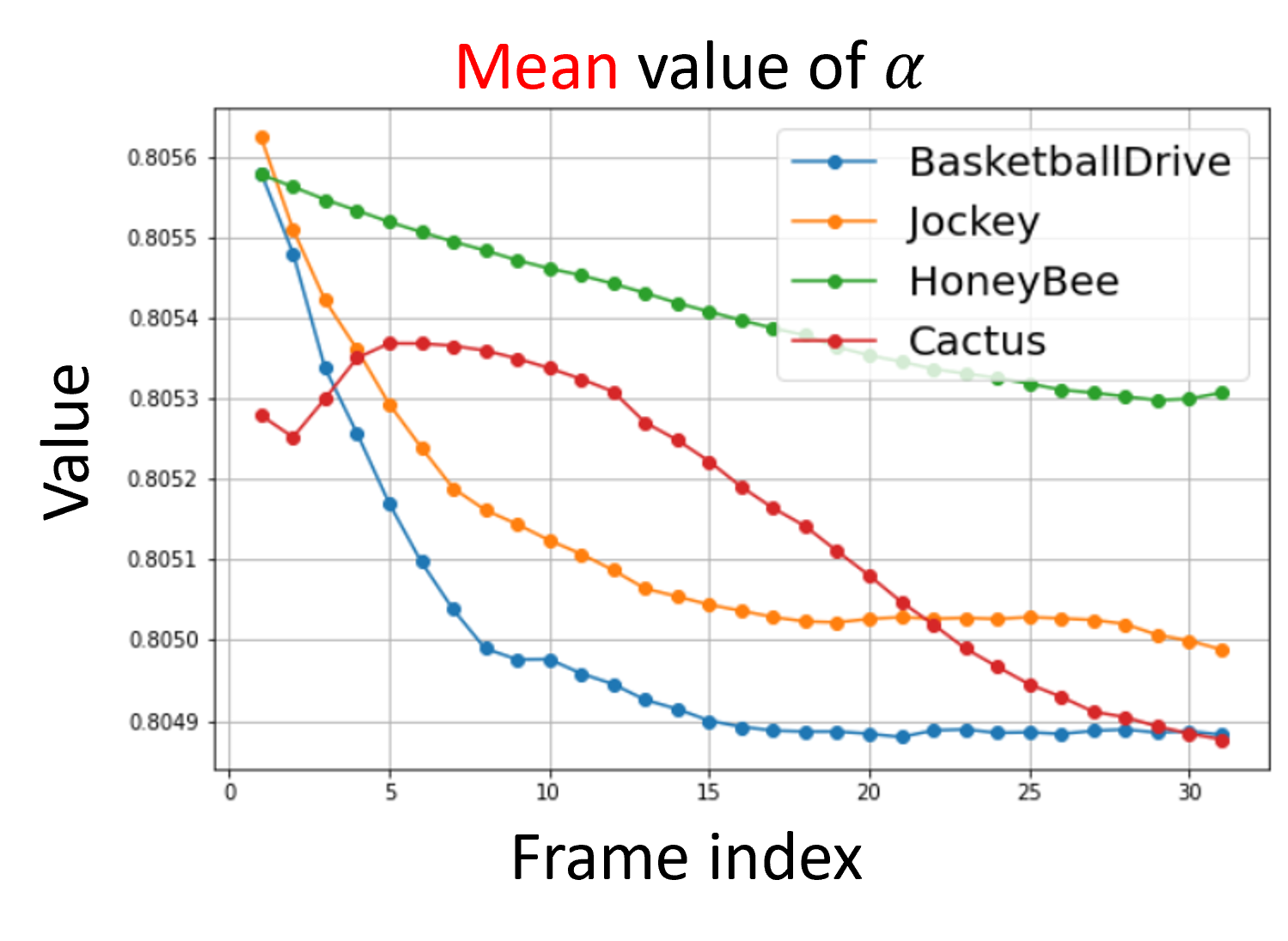}
  }
  
  \subfigure[Comparison of the minimum values.]{
    \includegraphics[width=0.4\linewidth]{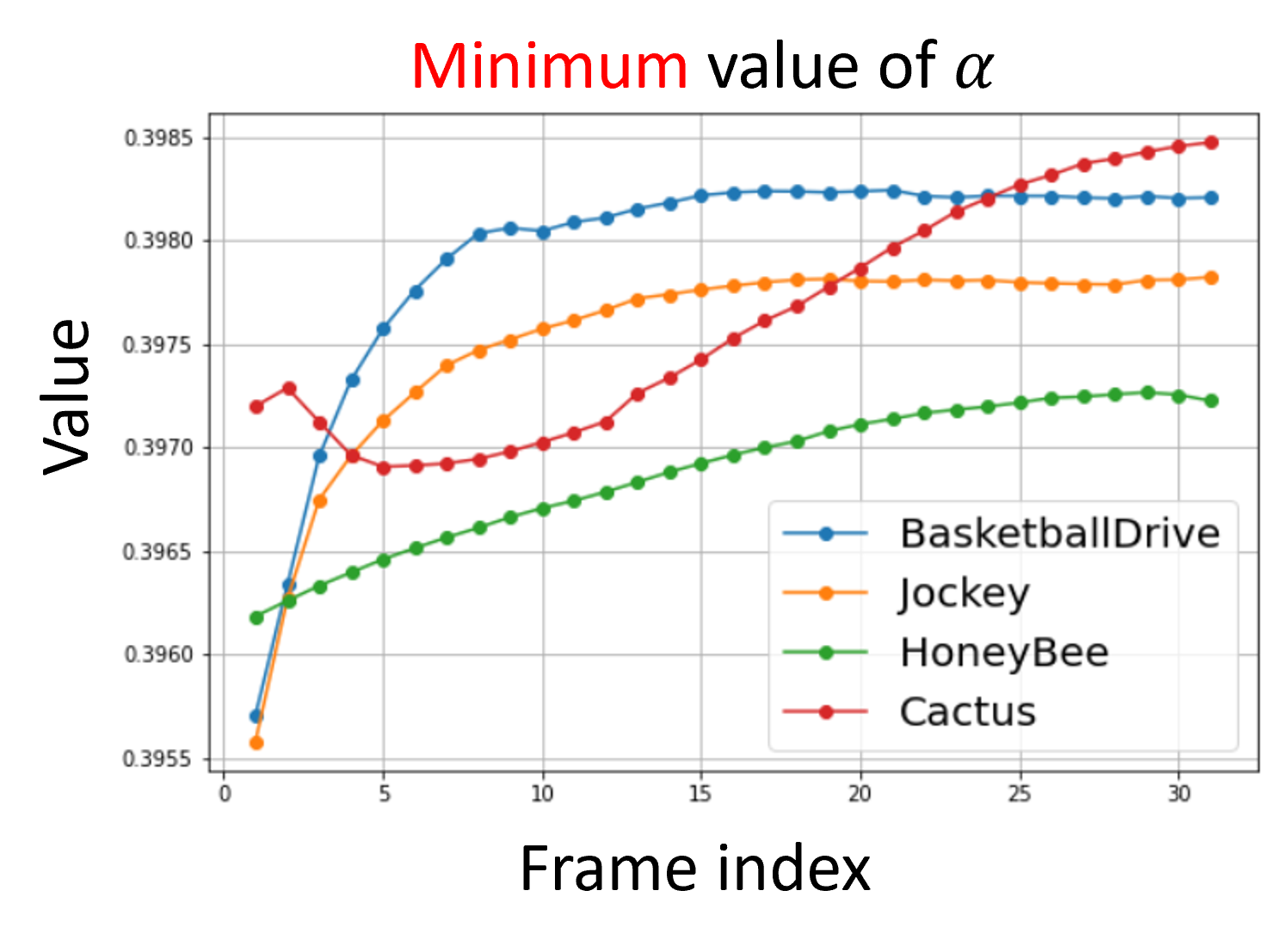}
  }
  
  \caption{Visualization of the maximum, mean, and minimum values of $\alpha$ during the coding of different sequences.}
  \label{fig:alpha_through_time}
\end{figure*}

Surprisingly, the results in Table~\ref{tab:ablation_state_signal} indicate that there is no significant difference in the coding performance when the temporal propagated optical flow maps are changed. This finding suggests that our feature map modulation did not fully achieve content adaptivity across different video content, as originally intended. Further insights into this are presented in the visualization of $\alpha$ values during coding, as shown in Figure~\ref{fig:alpha_through_time}. We selected the maximum, mean, and minimum values of $\alpha$ for four different sequences (\textit{BasketballDrive}, \textit{Jockey}, \textit{HoneyBee}, and \textit{Cactus}) to demonstrate that these values remain consistently close among the sequences. This observation underscores that our feature map modulation appears to enable the codec to learn a relatively fixed set of parameters $\alpha$ and $\beta$ to incorporate with modulated loss, rather than adapting them based on the video content, such as the temporal-propagated optical flow maps mentioned in Section~\ref{ssec:drift_error}.

\subsubsection{Complexity Analysis}
\label{ssec:complexity}
Table~\ref{tab:complexity} provides a comparison of the model complexity for each improvement introduced to the baseline framework. We analyze the complexity in three aspects: model size, number of operations, and the minimum buffer requirement for encoding a P-frame. The complexity changes are attributed to three factors: replacing the original MCNet with multi-scale MCNet, introducing parameters for feature map modulation, and including the quadtree partition-based entropy model. It is important to note that the increase in the number of operations is primarily due to the multi-scale MCNet, indicating that our improvements over the baseline model do not significantly impact model complexity.

\begin{table*}[t]
    \caption{Complexity analysis: Model Size (M) represents the number of model parameters in millions. KMACs/pixel denotes the number of Kilo Multiply-Accumulate operations per pixel, evaluated on 1080p (1920$\times$1080) sequences. Buffer Size indicates the number of full resolution feature maps (FRFM) required for P-frame coding.}
    \centering
    \small
    \begin{tabular}{c|ccc}
        \toprule
        \multirow{2}{*}{Methods}         & Model Size     & \multirow{2}{*}{KMACs/pixel} & Buffer Size  \\
                                         &  (M)           &                              & (FRFM)\\
        \hline
        Ho~\textit{et al.}~\cite{canfvc} & 31.0           & 2453.7                       & 13 \\
        + Multi-scale MCNet              & 37.7           & 2826.7                       & 13 \\
        + Feature Map Modulation         & 38.5           & 2862.9                       & 15 \\
        \multirow{1}{*}{} + Quadtree Partition-Based  & \multirow{2}{*}{42.0} & \multirow{2}{*}{2905.9} & \multirow{2}{*}{15} \\
        \multirow{1}{*}{} Entropy Model  &                &                              & \\
        \hline\hline
        DCVC~\cite{dcvc}                 &  8.0           & 1162.2                       & 3        \\
        DCVC-TCM~\cite{tcm}              & 10.7           & 1398.5                       & 67       \\
        DCVC-HEM~\cite{acmmm22}          & 16.8           & 1591.4                       & 68       \\
        DCVC-DC~\cite{li2023neural}       & 19.8           & 1274.1                       & 52       \\
        
        \bottomrule
    \end{tabular}
    \label{tab:complexity}
\end{table*}

The analysis shows that the model complexity increases gradually with each improvement. The introduction of the multi-scale MCNet leads to a slightly larger model size and increased computational requirements, as reflected in the number of operations per pixel. With the addition of feature map modulation, the model size further increases, and there is a minor increase in the number of operations per pixel. Finally, incorporating the quadtree partition-based entropy model results in a slightly larger model size and a marginal increase in the number of operations per pixel. However, the overall impact on model complexity remains manageable, as evidenced by the relatively small changes in model size and the minimal buffer requirement for P-frame coding.

We have also conducted a comparison of model complexity with recent works, namely DCVC~\cite{dcvc}, DCVC-TCM~\cite{tcm}, DCVC-HEM~\cite{acmmm22}, and DCVC-DC~\cite{li2023neural}. For the evaluation of model size, we specifically focus on the P-frame codec size as outlined in Table~\ref{tab:complexity}. In order to ensure a fair comparison with both the baseline framework and our CANF-VC++, we utilize their respective testing software.

When calculating the buffer size requirements for the mentioned models, we found that DCVC~\cite{dcvc} necessitates only a 3-channel reconstructed frame for reference. DCVC-TCM~\cite{tcm}, on the other hand, requires the buffering of a 64-channel full-resolution feature map (FRFM) along with a 3-channel reconstructed frame for reference. This results in a total buffer size equivalent to 67 FRFM.

In the case of DCVC-HEM~\cite{acmmm22}, similar to DCVC-TCM~\cite{tcm}, it also demands the buffering of a 64-channel full-resolution feature map and a 3-channel reconstructed frame for reference. Additionally, it requires buffering of latents for both motion coding and inter-frame coding, which leads to an additional requirement of 1 FRFM compared to DCVC-TCM~\cite{tcm}.

Lastly, for the evaluation of buffer size requirements in DCVC-DC~\cite{li2023neural}, it mandates the buffering of a 48-channel full-resolution feature map, a 3-channel reconstructed frame, and 2 latents for both motion coding and inter-frame coding, resulting in a cumulative total of 52 FRFM. In contrast to these recent advancements, our CANF-VC++ demonstrates a clear advantage by requiring a maximum of only 15 FRFM.

In summary, our proposed enhancements to the baseline framework provide improved rate-distortion performance without significantly increasing the model complexity. The incremental changes in model size, number of operations, and buffer requirements demonstrate the feasibility of integrating these techniques into practical video compression systems.

\subsection{Command Lines for HM and VTM}
For the encoding with HM~\cite{HM}, given an uncompressed video ``input.yuv'' of size W $\times$ H, we use the \emph{encoder\_lowdelay\_P\_main.cfg} configuration file with the following parameters: InputFile=input.yuv, FrameRate=FR, SourceWidth=W, SourceHeight=H, FramesToBeEncoded=N, IntraPeriod=32, GOPSize=8, DecodingRefreshType=2, and QP=Q, where FR, N, Q represent the frame rate, the number of frames to be encoded, and the quantization parameter, respectively. Q is set to 17, 22, 24, 27, 32.
For the encoding with VTM~\cite{VVCSoftware_VTM}, given an uncompressed video as "input.yuv" of size H x W, we use the \emph{encoder\_lowdelay\_P\_vtm.cfg} configuration from the official website with the following parameters: InputFile=input.yuv, FrameRate=FR, SourceWidth=W, SourceHeight=H, FramesToBeEncodec=N, IntraPeriod=-1, GOPSize=8, DecodingRefreshType=0, and QP=Q, where FR, N, Q represent the frame rate, the number of frames to be encoded, and the quantization parameter, respectively. Q is set to 17, 22, 24, 27, 32.

\section{Conclusion}
\label{sec:conclusion}

In this work, we present CANF-VC++, an updated learned video compression framework inspired by Ho~\textit{et al.}~\cite{canfvc}. Our goal was to rejuvenate and enhance the outdated video compression system by incorporating recent advancements in the field. We identified several limitations in the original model's training process, including conditioning signal generation, drift error handling, entropy coding efficiency, and training/testing misalignment. To address these issues, we proposed innovative solutions.

CANF-VC++ showcases substantial improvements in compression performance across popular datasets, such as UVG~\cite{uvg}, HEVC Class B~\cite{hevcctc}, and MCL-JCV~\cite{mcl}. Notably, it even outperforms the H.266 reference software VTM in terms of PSNR-RGB compression. Our experiments revealed that solely incorporating modulated loss, as done in previous work, was insufficient for significant performance gains. Consequently, we introduced feature map modulation as a complementary technique, resulting in remarkable improvements.

Although our feature map modulation did not fully achieve content adaptivity across different video content as initially intended, it proved to be a vital component when combined with modulated loss. Additionally, our research focused on enhancing the entropy model. We introduced an entropy model that considers both spatial and channel dependencies, leading to a notable reduction in required bit-rates and improved overall compression efficiency.

Overall, our work successfully applies the latest research advancements in video compression to the baseline framework, revitalizing the outdated system without significantly increasing model complexity. We hope our findings serve as inspiration for future video compression research, encouraging the systematic integration of existing tools and facilitating the adoption of new technologies.

{
\small
\bibliographystyle{ieee_fullname}
\bibliography{egbib}
}

\end{document}